\documentclass[]{aa}

\usepackage[dvips]{graphicx}
\usepackage{lscape,aalongtable}
\usepackage{supertabular}
\usepackage{amsmath}
\usepackage{amssymb}

\begin{document}

\title{Infrared study of the Southern Galactic star forming region associated 
with IRAS 14416-5937
}

\author{S. Vig\thanks{\emph{Present address:} INAF-Osservatorio
Astrofisico di Arcetri, Largo E. Fermi, 5 I-50125 Firenze, Italy}, S.K. Ghosh, D.K. Ojha \and R. P. Verma}

\offprints{S. Vig, \email{sarita@arcetri.astro.it}}

\institute{Tata Institute of Fundamental Research, Mumbai 400 005, India\\
}

\date{Received xxx / Accepted yyy}

\titlerunning{Infrared study of IRAS 14416-5937}
\authorrunning{S. Vig, et al.}
\abstract{}
{We have carried out an infrared study of the southern Galactic massive star
forming region associated with IRAS 14416-5937.
}
{This star forming region has been mapped simultaneously in two far infrared 
bands at $\sim 150$ \& 210 $\mu$m using the TIFR 1-m balloon borne telescope 
with $\sim 1'$ angular resolution. 
We have used 2MASS JHK$_s$ as well as Spitzer-GLIMPSE data of this region to 
study the stellar populations of the embedded young cluster. This region 
comprises of two sources, designated as A \& B and separated by $\sim2$ pc. 
The spectrum of a region located close to the source A obtained using the Long 
Wavelength Spectrometer (LWS) 
on-board the Infrared Space Observatory (ISO), is presented.  Emission
from warm dust and from Unidentified Infrared Bands (UIBs) is 
estimated using the mid-infrared data of the MSX survey.
}
{The spatial distributions of (1) the temperature of cool 
dust and (2) optical depth at 200 $\mu$m have been obtained
taking advantage of the similar beams in both the TIFR bands. A number of atomic
fine structure lines have been detected in the ISO-LWS spectrum, which have 
been used to estimate the electron density and the effective temperature 
of the ionising radiation in this region. From the near and mid 
infrared images, we identify a dust lane due north-west of source A. The dust 
lane is populated by Class I type sources. Class II type 
sources are found further along the dust lane as well as below it. 
Self consistent radiative transfer models of the two sources (A and B) are in 
good agreement with the observed spectral energy distributions. 
}
{The spatial distribution of young stellar objects in and around the dust lane 
suggests that active star formation is taking place along the dust lane and is 
possibly triggered by the expanding HII regions of A and B.}


\maketitle

\section{Introduction}

The southern Galactic high mass star forming region associated with IRAS 
14416-5937 is located at a distance of 2.8 kpc (Busfield et al., \cite{Bu06}). It 
corresponds to the radio source G316.8-0.1. A number of masers and molecular 
lines have been detected close to this star forming region. Both OH (Caswell
 \& Haynes, \cite{Ca87}) and H$_2$O (Caswell et al., \cite{Ca89}) masers have 
been observed here. Methanol (CH$_3$OH) maser (Caswell et al., \cite{Ca95}) 
showing variability has also been detected here. 
NH$_3$ (Vilas-Boas et al., \cite{Vi00}), CI (Huang et al., \cite{Hu99}), CO 
(Whiteoak et al., \cite{Wh82}; White \& Phillips, \cite{Wh83}), CS (Bronfman 
et al., \cite{Br96}; Juvela, \cite{Ju96}) and H$_2$CO (Gardner \& Whiteoak, 
\cite{Ga84}) lines have been detected in this star forming region. Walsh et al.
 (\cite{Wa98}) present high angular resolution ($\sim1.5''$) radio continuum 
and methanol maser data at 6.67 GHz for this source as a part of their survey. 
The 
location of IRAS 14416-5937 being in the extreme southern sky, no detailed 
study of this region exists in the literature. In this paper, we have carried 
out a systematic study of the star forming region associated with IRAS 
14416-5937.

IRAS 14416-5937 has been 
studied in the infrared wavebands with the aim of investigating the emission 
from dust, the dust temperature, the energetics and the stellar populations of 
the associated cluster in this region. In Section 2, we present the 
far infrared observations and a description of other available data used in 
this paper. Section 3 describes the results and Section 4 deals with the 
radiative transfer modelling carried out. In Section 5, we discuss all the 
results, and a brief summary is presented in Section 6.

\section{Observations and Data Reduction}

\subsection{Far infrared observations}
The Galactic star forming region associated with IRAS 14416-5937 has been 
observed using two-band far infrared (FIR) photometer
system at the Cassegrain focus of
the TIFR 100 cm (f/8) balloon borne telescope. The observations were carried
out during the balloon flight from the TIFR Balloon Facility, Hyderabad in
India (latitude $17^o.47$ North, longitude $78^0.57$ East) on Feb 20, 1994.
Details of the telescope and the observational procedure are given by Ghosh et
 al (\cite{Gh88}). The two FIR bands use a pair of 2$\times$3 composite 
silicon bolometer arrays, cooled to 0.3K by liquid $^3$He, which view
identical parts of the sky simultaneously. The field of view corresponding to 
each detector is $1.6'$. The sky is chopped along the cross-elevation axis at 
10 Hz with a throw of $4.2'$.  
The bandpasses of the two bands are defined by sets of cool filters. 
The spectral response was measured with a Michelson interferometer using a 
Golay cell as a comparison detector. The effective wavelengths of the two FIR 
bands are 150 and 210 $\mu$m  respectively. These effective wavelengths  
correspond to a source spectrum of 30 K
gray body with an emissivity dependence of 
$\epsilon_{\lambda}\propto\lambda^{-2}$.
The absolute positions were calibrated using the 
observations of catalogued stars with the optical photometer located at the 
focal plane of the telescope. The field of view of this optical photometer is 
offset with respect to the infrared field.
The planet Jupiter was observed for absolute flux calibration as well as
for the determination of the instrumental Point Spread Function (PSF).

The simultaneous mapping in the two FIR bands was carried out by 
raster-scanning the telescope along the cross-elevation axis across the 
target area under study and stepping along elevation at both
the ends of the scans. A region $\sim 32\arcmin \times 20\arcmin$
centred around IRAS 14416-5937, was mapped. 
The FIR signals were gridded 
into a two-dimensional (elevation $\times$ cross-elevation) matrix with a 
pixel size of $0.3\arcmin \times0.3\arcmin$. The observed chopped signal 
matrix was deconvolved using the Maximum Entropy Method similar
to that of Gull \& Daniell (\cite{Gu78}) (see Ghosh et al, \cite{Gh88} for 
details). An angular resolution of $\sim 1\arcmin$ has been achieved in the FIR 
maps using this method. The estimated error on absolute flux densities 
for the TIFR bands is $\sim$ 10\%, primarily from the calibration uncertainties 
using the planet.  

\subsection{Other available datasets}

\subsubsection{IRAS-HIRES}

The Infrared Astronomical Satellite (IRAS) survey data at all the four 
wavelength bands for the region around IRAS 14416-5937 were HIRES-processed 
 at the Infrared Processing and Analysis Center (IPAC), Caltech. 
HIRES processing employs the Maximum Correlation Method (MCM; Aumann 
et al., \cite{Au90}) to construct (resolution enhanced) coadded images.
These maps have been used in the present study to quantify the flux 
densities and angular sizes at the four infrared bands. They have also been 
used to generate the temperature and optical depth maps of interstellar dust. 
An upper limit on error in flux density in each band has been estimated
 by integrating several circular regions ($3'$ diameter) with no point-like 
source, in the local neighbourhood of the  respective maps.

\subsubsection{MSX}

The Midcourse Space Experiment (MSX) was a satellite experiment which surveyed 
the entire Galactic plane within $|$b$|\le 5^{\circ}$ in four mid infrared 
wavebands: 8.3, 12.1, 14.7 and 21.3 $\mu$m with a spatial resolution of 
$\sim18\arcsec.3$ (in all four bands) during 1996 - 1997 (Price et 
al., \cite{Pr01}). The infrared instrument on MSX, designated SPIRIT 
III, was a 35 cm clear aperture off-axis telescope with focal plane arrays. 
In the present study, we have used the panoramic 
images of the region around IRAS 14416-5937 to extract sources and obtain the 
integrated flux densities for constructing the spectral energy distribution 
(SED) of IRAS 14416-5937. Upper limits to errors on flux densities 
have been estimated using a procedure similar to that for IRAS-HIRES (see 
section 2.2.1).

\subsubsection{2MASS}

The Two Micron All Sky Survey (2MASS) used two highly-automated 1.3-m 
telescopes, one at Mt. Hopkins, Arizona (USA), and one at Cerro Tololo 
Inter-American Observatory (CTIO), Chile, to uniformly 
scan the entire sky in three near-infrared bands: J (1.25 $\mu$m), H (1.65 
$\mu$m), and Ks (2.17 $\mu$m), using a pixel size of $2.0''$. The survey was 
completed in 2001. 
We have used the point sources from the region around 
IRAS 14416-5937 from 2MASS Point Source Catalog (PSC) in this 
study.  
The 2MASS PSC is complete down to J $\le15.8$, H $\le15.1$ and K$_s\le14.3$
magnitudes for S/N$>10$, in the absence of confusion. The J, H and
K$_s$ magnitudes of the extracted sources have been used to make
colour-magnitude and colour-colour diagrams to study the
embedded cluster in this region. The JHK$_s$ magnitudes and images were 
taken from IPAC. 

\subsubsection{Spitzer-GLIMPSE}

The Spitzer Space Telescope (Werner et al., \cite{We04}) was launched 
in space in August 2003 and 
consists of a 0.85-meter telescope with three cryogenically cooled instruments:
 InfraRed Array Camera (IRAC), InfraRed Spectrograph and Multiband Imaging 
Photometer for Spitzer. IRAC is a four-channel camera that provides 
simultaneous $5.2'\times5.2'$ images at 3.6, 4.5, 5.8, and 8 $\mu$m with a 
pixel size of $1.2''\times1.2''$ (Fazio et al., \cite{Fa04}). In the GLIMPSE 
(Galactic Legacy Infrared Midplane Survey Extraordinaire; Benjamin et al., 
\cite{Be03}) project, Spitzer Space Telescope is surveying approximately 220 
square degrees of the Galactic plane covering a latitude range of 
$|b|<1^{\circ}$ and a longitude range of $10^{\circ}\le l\le65^{\circ}$, 
$-65^{\circ}\le l\le-10^{\circ}$. This survey is carried out in the 4 
IRAC bands. The catalog lists sources within each surveyed
$2^{\circ} \times2^{\circ}$ region. The sources around IRAS 14416-5937 
have been extracted from the GLIMPSE More Complete Archive. 
The GLIMPSE archive 
catalogs contain point sources with peak signal-to-noise ratio greater than 5 
in at least one band. The magnitudes of the extracted sources have been used 
in making the colour-magnitude and colour-colour diagrams. The 
Spitzer-GLIMPSE images have been obtained using the software `Leopard'. These 
images have been used to study the spatial distribution of sources as well as 
mid infrared emission from this region.

\subsubsection{SUMSS}

The Sydney University Molonglo Sky Survey (SUMSS) is a radio imaging survey of
the southern sky ($\delta < -30^{\circ}$; Bock et al., \cite{Bo99}). 
This survey, using the Molonglo Observatory Synthesis Telescope (MOST), 
is being carried out at 843 MHz. The MOST consists of two 
cylindrical paraboloids, 778m x 12m, 
separated by 15m and aligned East-West. 
The radio image of 
IRAS 14416-5937, extracted from the SUMSS Archive, has been used to study the 
distribution of ionised gas around this region. The synthesized beam size is 
$43''\times50''$. 

\subsubsection{ISO}

The Infrared Space Observatory (ISO) was an astronomical satellite 
experiment consisting of a telescope with primary mirror of size 60 cm and 
operational between 1995 and 1998. The various scientific instruments on-board 
the ISO operated between 2.5 - 240 $\mu$m.
We have used the data from the Long Wavelength Spectrometer
(LWS, Clegg et al., \cite{Cl96}) for a region close to IRAS 14416-5937 
between $43-197$ $\mu$m.
The LWS spectrum is for a region centred at $\alpha_{2000}$ = $14^h$ $45^m$ 
$21.0^s$, $\delta_{2000}$ = -59$^{\circ}$ 48$'$ 14$''$. The version of the ISO 
data used in this paper corresponds to the Highly Processed Data 
Product (HPDP) 
sets called `Uniformly processed LWS L01 spectra' by Lloyd et al. (\cite{Ll03}),
obtained from the ISO Data Archive.

\section{Results}

\subsection{Emission from dust}
\begin {figure*}
\hskip -1cm
\includegraphics[height=7.0cm]{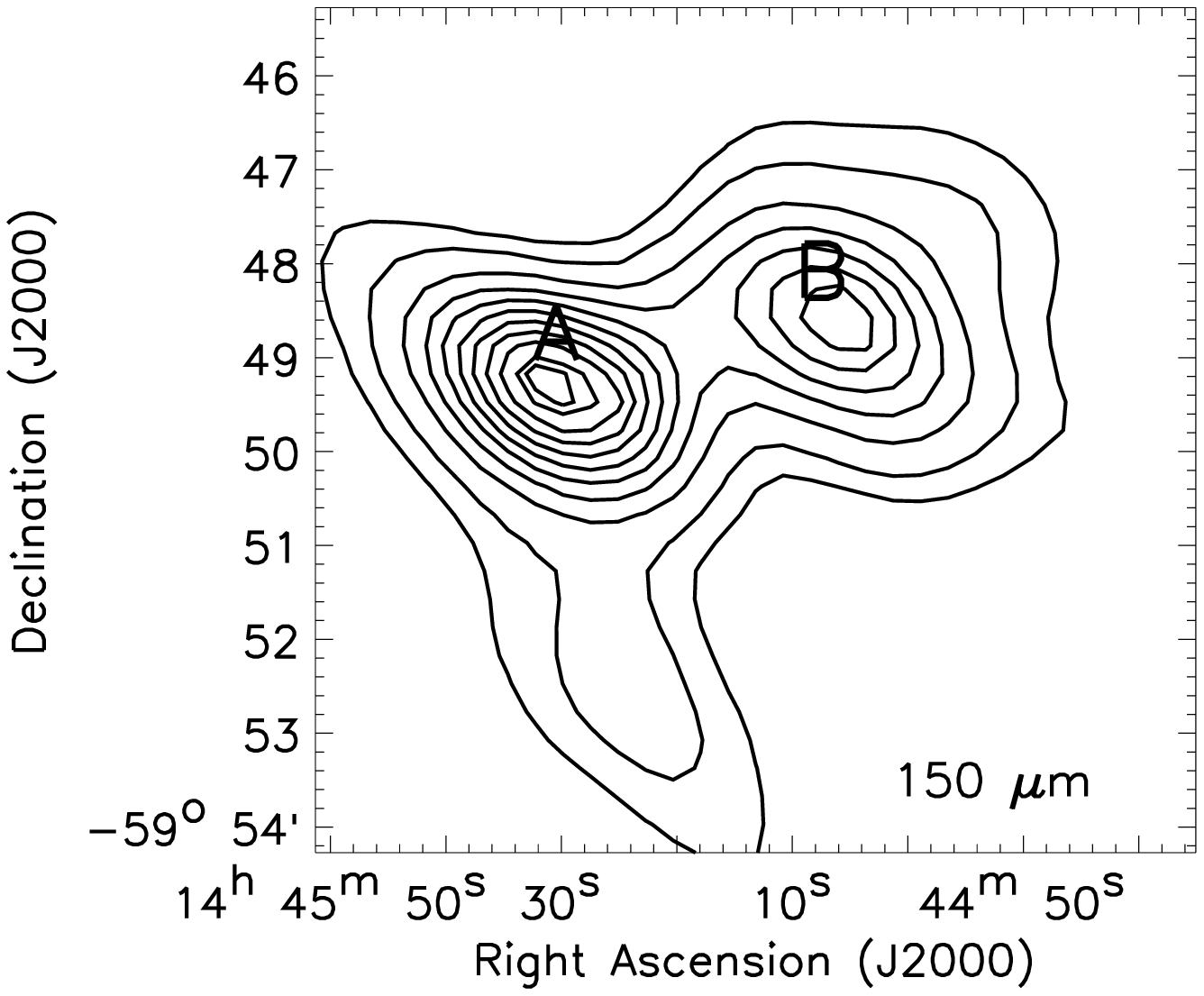}
\hskip -2cm
\includegraphics[height=7.0cm]{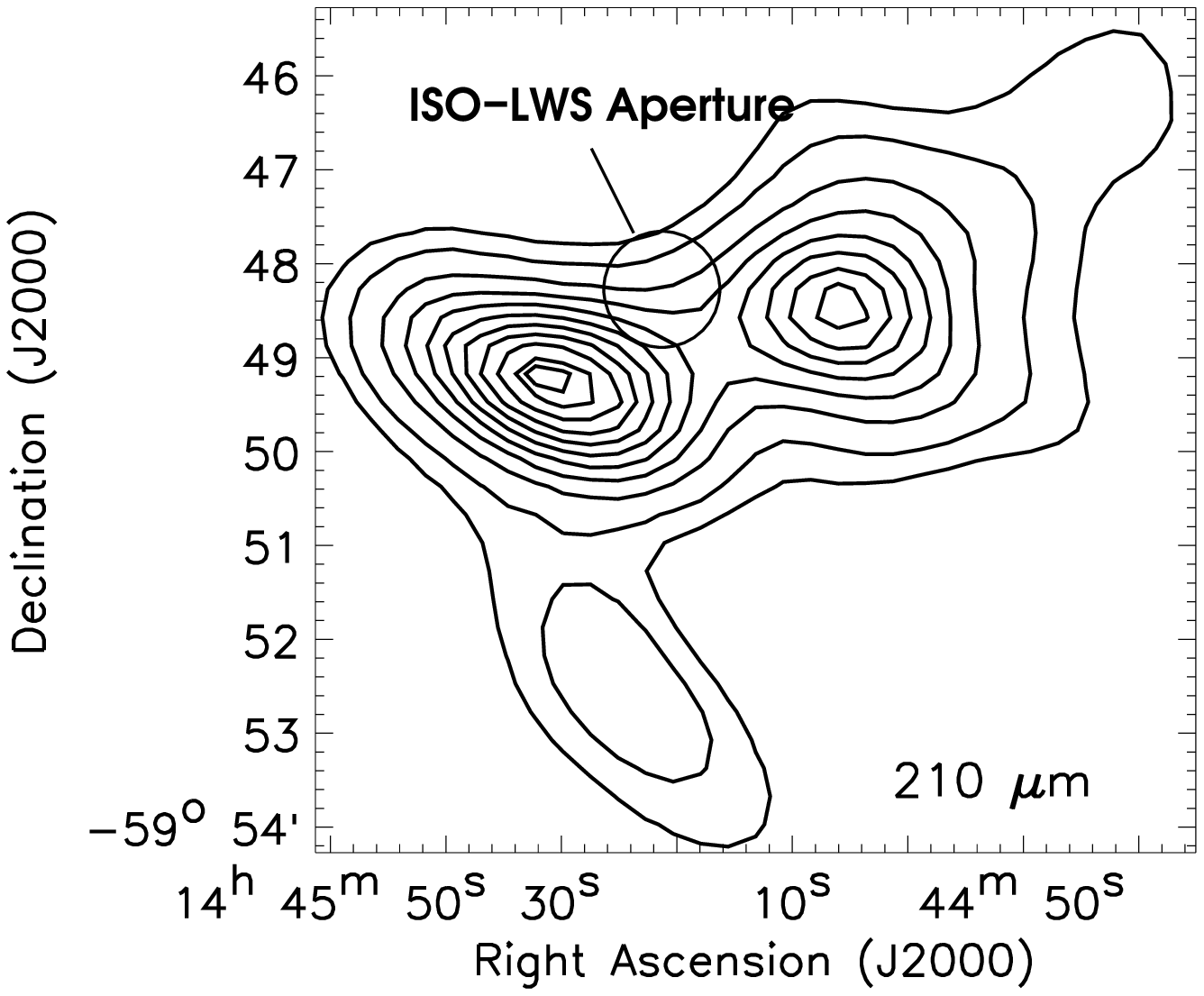}
\caption{The intensity map for the region around IRAS 14416-5937 at 150 $\mu$m
(left) and 210 $\mu$m (right). Contour levels are at 5, 10, 20, 30, 40,
50, 60, 70, 80, 90, 95\% of peak intensity of 2450 Jy/sq arc min (left) and
1367 Jy/sq arc min (right).
}
\label{FIRmap}
\end {figure*}

\begin {figure*}
\includegraphics[height=6.0cm]{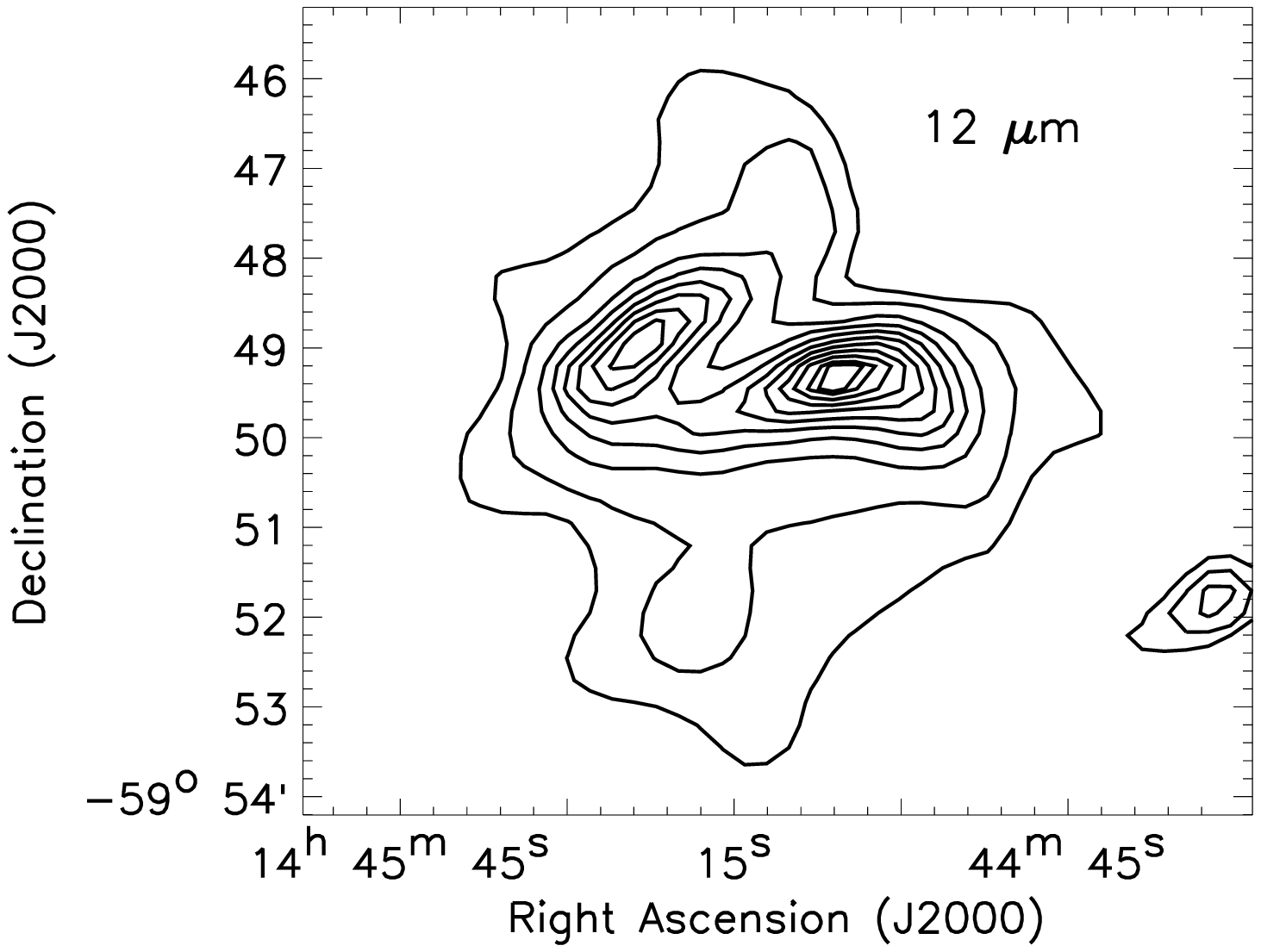}
\includegraphics[height=6.0cm]{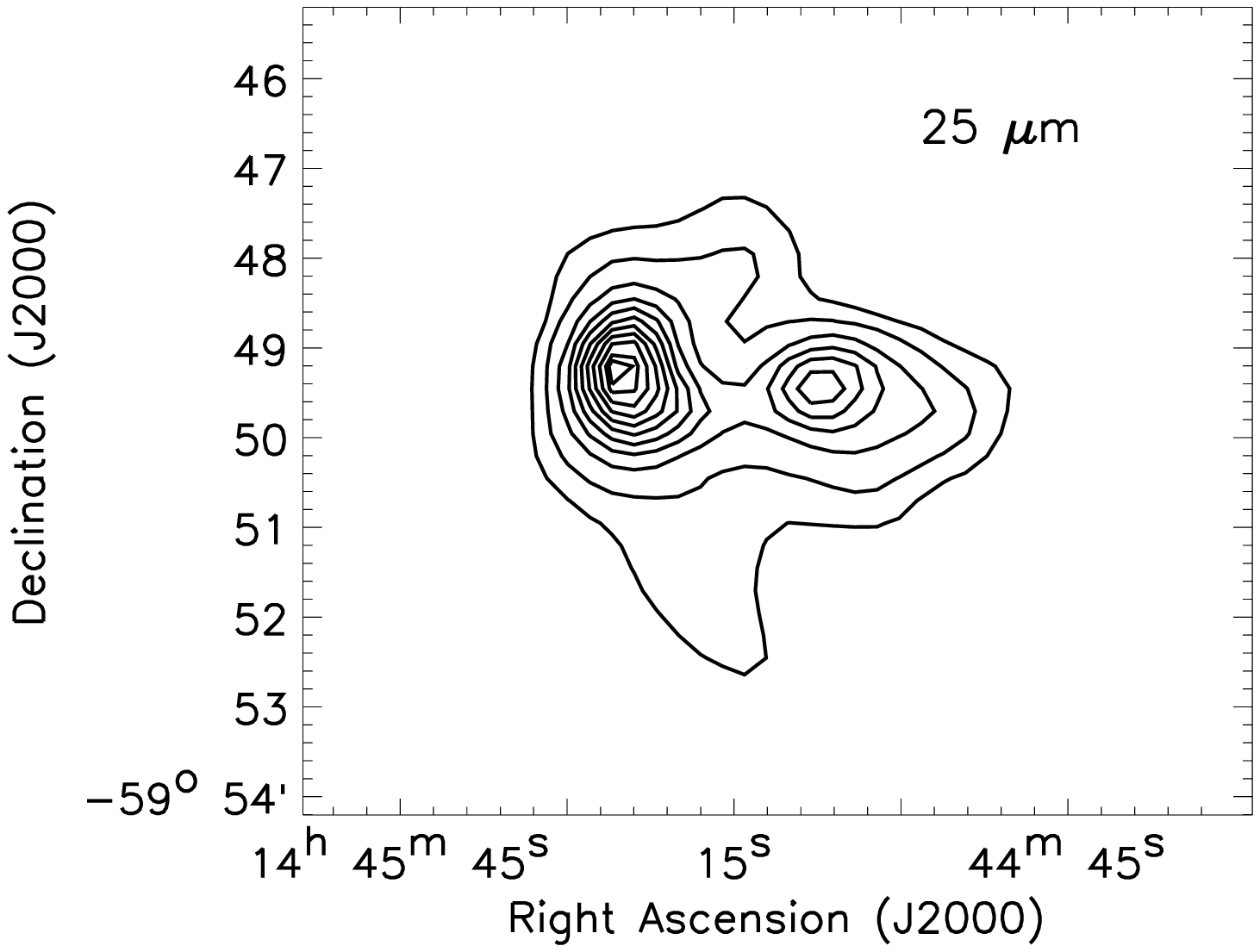}
\includegraphics[height=6.0cm]{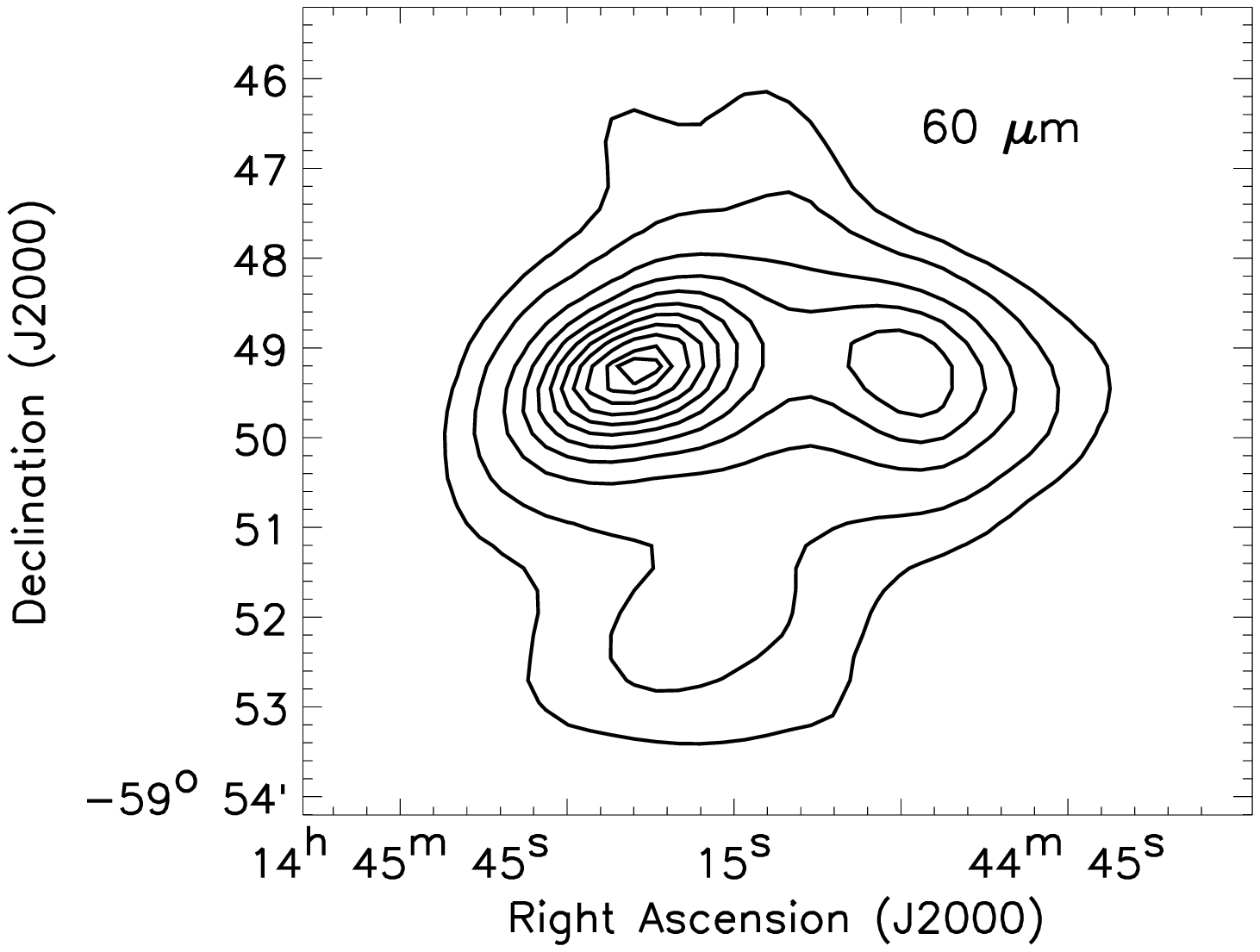}
\includegraphics[height=6.0cm]{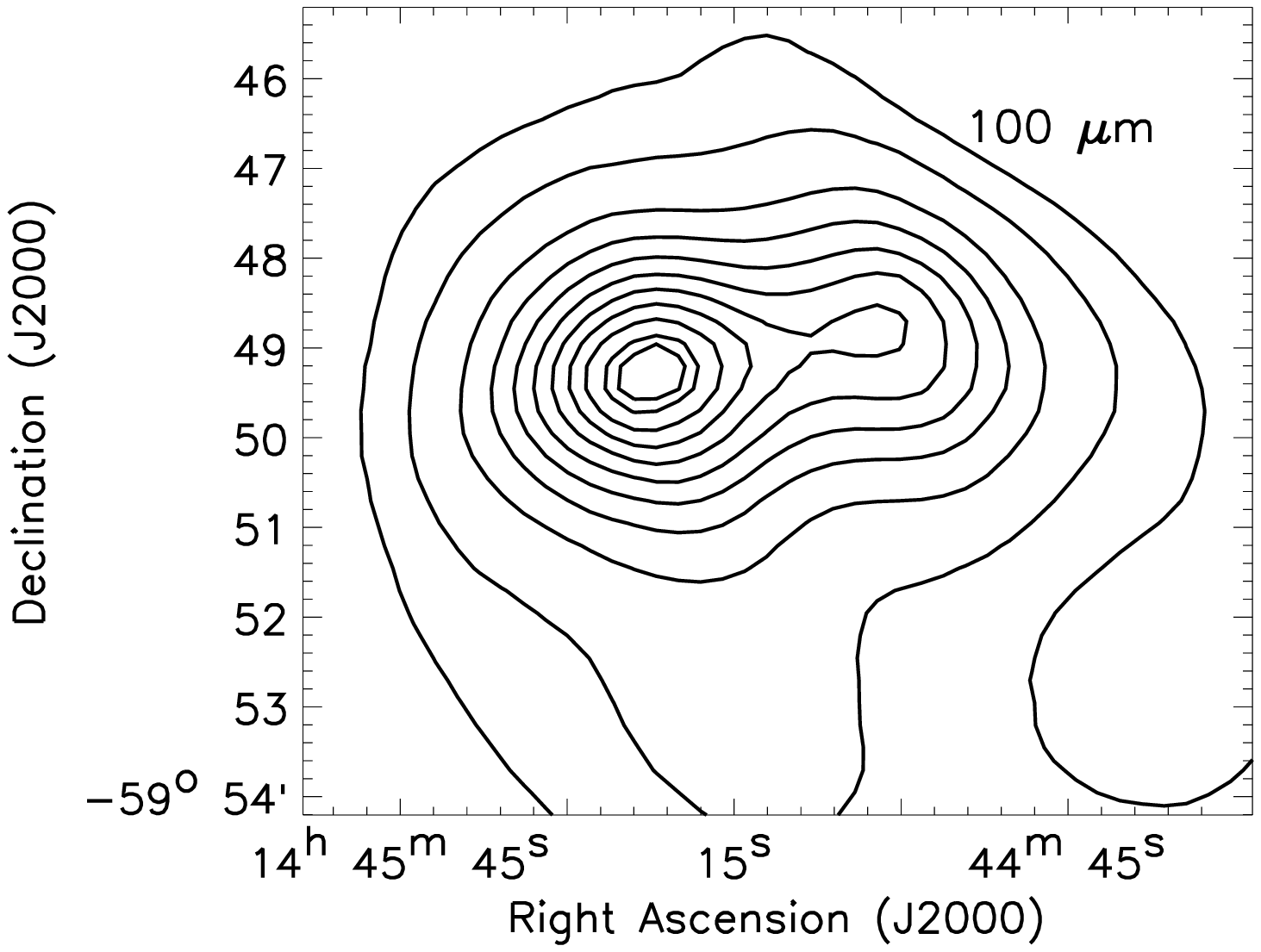}
\caption{The IRAS-HIRES intensity map for the region covering IRAS 14416-5937
at 12 $\mu$m (top left), 25 $\mu$m (top right), 60 $\mu$m (bottom
left) and 100 $\mu$m (bottom right). The contours are at 5, 10, 20, 30, 40, 
50, 60, 70, 80, 90 and 95 \% of the peak value of
155 Jy arcmin$^{-2}$, 1380 Jy arcmin$^{-2}$, 3360 Jy arcmin$^{-2}$ and 2410 Jy
arcmin$^{-2}$ at 12, 25, 60 and 100 $\mu$m, respectively.
}
\label{HIRES}
\end {figure*}

The thermal emission from dust in the TIFR-bands at 150 and 210 $\mu$m is shown in 
Figure \ref{FIRmap}. These deconvolved maps show emission from two sources 
designated A (east) and B (west).  The dynamic range of the maps is quite good 
and contours 
are shown upto 5\% level of the respective peak intensities (2450 Jy/sq arc min and
1367 Jy/sq arc min at 150 and 210 $\mu$m, respectively). The HIRES-processed 
maps at all the four IRAS bands (12, 25, 60 and 100 $\mu$m) are shown in 
Figure \ref{HIRES}. 
Similar to the TIFR maps, both the peaks A and B are seen clearly in 
the 12, 25 \& 60 $\mu$m maps but only a hint of B appears at 
100 $\mu$m.
The peak intensities in the HIRES maps correspond to 155, 1380, 3360 and 
2410 Jy/sq arc min at 12, 25, 60 and 100 $\mu$m respectively. 
The flux densities, obtained by integrating circular regions of diameter 
3$'$ centred on peaks A and B from the TIFR, IRAS-HIRES and MSX maps, are 
listed in Table \ref{fluxes}. 
The IRAS PSC lists a single source in this region. The corresponding flux 
densities and 
position of the IRAS PSC source are also listed in the table.  

Although the IRAS-HIRES maps have a much higher dynamic range, the angular 
resolution of TIFR maps are superior to the IRAS maps (at least at 60 and 100 
$\mu$m) because of the smaller and circular beams employed. Since the TIFR 
beams are nearly identical at both the FIR bands and all the observations are 
simultaneous, this data set is useful in mapping the colour 
temperature and dust optical depth with good angular resolution. 
The flux density, $F_{\nu}$, for optically thin emission 
can be written as 
$$F_{\nu} = \Omega B_{\nu}(T_d)\tau_{\nu}$$
where, $\Omega$ is the solid angle of the region under consideration,  
$B_{\nu}$ is the Planck function and $T_d$ is the dust temperature. Assuming 
$\tau_{\nu} \propto \nu^\beta$, it can be shown that the ratio of flux 
densities at any two wavelengths is a function of $T_d$ and $\beta$. For 
various dust temperatures ($T_d$) and an assumed value of $\beta$, a look-up 
table is generated for the ratio of flux densities. The colour temperature and 
optical depth maps have been generated using this interpolation table relating 
the ratio of signals 
detected for the two bands to the dust temperature for the assumed emissivity 
law, $\epsilon_{\lambda}\propto\lambda^{-2}$. It is important to note that the 
morphology of the contours of dust temperature and opacity are not sensitive 
to  the assumption $\beta$. Further details can be found in 
the Appendix A of Mookerjea et al. (\cite{Mo00}). The generated maps of colour 
temperature T(150/210) and dust optical depth at 200 $\mu$m ($\tau_{200}$),
are shown in Figure \ref{Ttaubal}. 

Using the emission in the MSX bands (8.3, 12.1, 14.7, 21.3 $\mu$m) 
for the region 
around IRAS 14416-5937, we have modelled the thermal continuum from  
interstellar dust alongwith emission in the Unidentified Infrared Bands 
(UIBs) following the scheme developed by Ghosh \& Ojha (\cite{Gh02}). 
In this scheme, the emission from  each pixel in the MSX images is a combination of two 
components: (i) thermal continuum from dust grains
(gray body) and (ii) the emission from the UIB features falling within the MSX 
band. The scheme assumes that dust emissivity follows the power
law of the form $\epsilon_{\lambda}\propto\lambda^{-1}$ and the total radiance
due to UIBs in the 12 $\mu$m band is proportional to that in the 8 $\mu$m
band. 
The dust emissivity law depends on wavelength. We have used  
emissivity laws, $\epsilon_{\lambda}\propto\lambda^{-1}$, for $\lambda<100$ 
$\mu$m and, $\epsilon_{\lambda}\propto\lambda^{-2}$, for $\lambda>100$ $\mu$m, 
which is generally used (Scoville \& Kwan, \cite{Sc76} ). 
A self consistent non-linear chi-square minimization technique is used to
estimate the total emission from the UIBs, dust temperature and optical depth
in mid infrared (10 $\mu$m). 
The spatial distribution of UIB emission is shown in Figure \ref{pah}.
The peak strength of the modelled UIB emission is $1.3\times10^{-4}$ W 
m$^{-2}$ Sr$^{-1}$, close to peak A. 

\subsection{Emission from gas}

The radio continuum emission from the region around IRAS 14416-5937 from SUMSS survey at 843 MHz
is shown in Figure \ref{SUMSS}. 
The rms noise in the map is $\sim6$ mJy/beam. The 
radio emission peaks at ($\alpha_{2000}$ = $14^h$ $45^m$ $23.52^s$, 
$\delta_{2000}$ = -59$^{\circ}$ 49$\arcmin$ 25.0$\arcsec$). The integrated 
radio flux density upto 5\% contour level is $\sim37.5$ Jy over 30 arcmin$^2$.

The ISO-LWS beam, centered at a location which is $\sim1.4'$ to the 
north-west of IRAS 14416-5937 - A, is shown in Figure \ref{FIRmap} 
(right). This spectrum, extending from $43-197$ 
$\mu$m is shown in Figure \ref{ISO_cloudy}. A
number of lines are prominently detected. The atomic fine structure
lines with good signal to noise ratio are identified in the Figure and their 
line fluxes are given in Table \ref{iso_tab1}. The spectrum is dominated by 
the fine structure lines of [N II], [N III], [O III], [C II] and 
[O I]. The line fluxes are extracted by fitting Gaussian functions and 
integrating the area under the curve after removing the underlying local 
continuum estimated through a polynomial fit to the baseline. 
The line fluxes, normalized to [C II] line at 158 $\mu$m, are also
presented in Table \ref{iso_tab1}.

\subsection{Embedded cluster}

To study the embedded cluster associated with IRAS 14416-5937, we have 
selected the 2MASS sources in a square region of size $\sim8.3'$ around 
the IRAS source. This square region includes the sources A and B as 
well as the dust 
lane seen in the near and mid infrared images (details in Section 5.3). 
A total of 1847 sources are present in the 2MASS catalogue, of which 
722 are detected in all JHK$_s$ bands with good quality flags 
(i.e. rdflg = 1 to 3). The stellar populations of this region have been 
investigated using the colour-magnitude (CM; J vs J - H) and colour-colour
(CC; J - H vs H - K) diagrams of this sample of 722 sources.
These are shown in Figure \ref{cmcc_2mass}. 
In the CM diagram, the nearly vertical solid lines from left to right represent
 the ZAMS curves (for a distance of 2.8 kpc) reddened by A$_V$ = 0 and 20
magnitudes, 
respectively. The slanting lines trace the reddening vectors of these main 
sequence stars. In the CC diagram, the locii of the main sequence and giant
branches are shown by the solid and dotted lines, respectively. The short-dash
line respresents the locus of T Tauri stars (Meyer et al., \cite{Me97}). The
three parallel dot-dash straight lines follow their reddening vectors. The
long-dash line represents the locus of Herbig Ae/Be stars (Lada \& Adams, 
\cite{La92}). We have assumed extinction values of A$_J$/A$_V$ = 0.282,
A$_H$/A$_V$ = 0.175 and A$_{K_s}$/A$_V$ = 0.112 from Rieke \& Lebofsky 
(\cite{Ri85}). All the 2MASS magnitudes as well as the curves are in the 
Bessel \& Brett (\cite{Be88}) system. In Figure \ref{cmcc_2mass}, the sources 
lying above the reddening curve of the ZAMS spectral type B0 are shown as 
asterisks while the stars with infrared 
excess are shown as open circles. These are sources to the right of 
the reddening vector drawn from the bottom of the main sequence stars or 
sources lying in the T tauri and HeAeBe zones in the CC diagram. The other 
sources are represented by dots. 
It is important to note that the CM and CC diagrams are useful tools 
for estimating the approximate nature of the stellar populations within the 
cluster in the absence of any spectroscopic data.

While the near infrared sources in this region have been studied using 2MASS, 
the near-to-mid infrared sources in this region have been extracted from the 
Spitzer-GLIMPSE catalogs. In a region identical to the 
one used for extracting the 2MASS sources, a total of 2087 sources were 
detected. 
For our analysis, those sources were selected which have flux 
calculation method flag (MF) equal to 0 (good quality). A total of 130  
sources (with MF=0) were detected in all the four IRAC bands.
All these 130 sources have been plotted in a colour-colour diagram ([3.6] - 
[4.5] vs [5.8] - [8.0]) which is shown in Figure \ref{cc_glim}. We have used 
the IRAC colour-colour diagram and the models of Allen et al. (\cite{Al04}) to 
identify the young stellar objects in this region. The solid-line square in the diagram approximately delineates the region occupied by the Class II sources 
whereas the dotted-line square covers the region occupied by the Class I 
sources as shown in the models of Allen et al. (\cite{Al04}) (see 
their Figure 4). There is a region of 
overlap of these two boxes and the sources in this region will be referred to 
as Class I/II sources. In our sample of 130 
sources, we find that 14, 11 and 11 sources can be classified as Class I, Class 
I/II and Class II, respectively. These are likely to be associated with 
IRAS 14416-5937 and to be evolving towards the main sequence. In Figure 
\ref{cc_glim}, the open circles denote Class I sources, the open squares 
represent Class I/II sources, the filled triangles denote Class II sources and 
the cross symbols represent the other sources.

\section{Radiative transfer modelling}

The modelling procedure described in this section has been used to interpret 
the results obtained.

\subsection{Continuum emission from dust and gas}

We have carried out radiative transfer modelling of the sources IRAS 14416-5937 A and B to obtain the various physical parameters of this region. 
The radiative transfer equations have been solved 
assuming a two-point boundary condition for a spherically symmetric cloud of 
dust and gas. The gas exists throughout the modelled cloud. The dust, 
however, exists in a spherical shell with a cavity at the center. The cavity 
represents the region where dust temperature would be higher than its 
sublimation temperature due to stronger radiation field. In 
the spherical shell where gas and dust co-exist, the gas-to-dust ratio is held 
constant. The cloud is heated internally by centrally embedded sources and
by an external radiation field due to the average Galactic interstellar
radiation field (ISRF).
The position of the ionisation front depends on the effective 
temperature and luminosity of the exciting star. 

For modelling the observed spectral energy 
distribution, two types of dust have been explored.
The first type of dust grains is from Draine \& Lee (\cite{Dr84}), 
hereafter referred to as DL type of grains. The physical properties of the 
grains, viz., absorption and scattering efficiencies, the scattering 
anisotropy factor for all sizes and frequencies used in the model were taken 
from the tables of B. T. Draine's homepage\footnote{http://www.astro.princeton.edu/$\sim$draine/dust/} which are computed in a similar 
manner to that by Laor 
\& Draine, (\cite{La93}). Three types of most commonly accepted variety of 
dust grains have been used here: (i) Graphite, (ii) Astronomical Silicate and 
(iii) Silicon Carbide. The second type of dust is from Mezger, Mathis 
\& Panagia (\cite{Me82}; hereafter MMP). This type of dust consists of 
graphite and 
silicate only, but their absorptive and scattering properties differ 
substantially from those for the DL case, particularly in mid infrared.
The relative abundances of the types of grains are used as parameters of our 
modelling. 

The cloud is parameterised by the following quantities: the geometry of cloud 
(outer size and inner size of dust cavity), total radial optical depth at a
specified wavelength, radial dust and gas density distribution laws ($r^0$, 
$r^{-1}$, $r^{-2}$). The gas to dust ratio is a parameter which is held 
constant throughout the cloud (except for the central dust-free cavity). The 
luminosity of the embedded source is obtained by integrating the area under 
the observed SED. The physical sizes of the cloud, the radial optical depth, 
dust composition and the radial density distribution are varied to obtain a 
good match to the observations. The observed angular sizes and luminosity have 
been used to constrain the model. With this scheme, a best fit model matching 
the observed SED and angular sizes at selected wavelengths and the radio 
continuum flux is obtained. Further details of the modelling scheme 
are given by Mookerjea \& Ghosh (\cite{Mo99}). 

The SEDs of both these sources (IRAS 14416-5937 A \& B) are constructed using 
the flux densities at the two TIFR bands, the four IRAS bands (from HIRES maps)
 and the four MSX bands. It may be noted that the peak emission associated with
 B shifts northward with increasing wavelength. This could be attributed to the
cold dust lane. Although the details of the geometry are unclear, we have 
considered the peak emission of B associated with MSX bands and integrated the 
flux density in a circle of diameter 3$'$ around this MSX peak in all the 
bands. Since A and B are separated by $\sim2.5'$, the flux densities in the 
slight overlap region are distributed in the ratio of the intensities of A and 
B peaks.

\subsubsection{IRAS 14416-5937 - A}

The total luminosity of this source is $1.4\times10^5$ L$_{\odot}$ for a 
distance of 2.8 kpc. The best fit radiative transfer model corresponds to a 
uniform density distribution of dust and gas. The DL type of dust fits the data 
better. The relative fraction of the two constituent grain types 
Si:Gr is 11:89 for the best fit model. The predicted spectrum by the 
best fit model has been compared with the observations in Figure \ref{radtran} (left) and 
the corresponding parameters obtained from this model are tabulated in Table 
\ref{radtable}. The cloud size (outer radius) is 3.4 pc and the radial optical depth at 100 
$\mu$m is 0.007. A single ZAMS star of spectral type O7-O6.5 has 
been used as the centrally exciting source. From the model, the radius of the 
ionised region is determined to be 0.5 pc. The radio flux density predicted 
by the model at 843 MHz is 2.6 Jy for a gas-to-dust ratio of 100. This is 
lower than the 
measured value of 10.2 Jy obtained by integrating within a circular region of 
radius 0.5 pc around the radio peak (see Figure \ref{SUMSS}). This 
could be due to either gas-to-dust ratio and/or due to clumpy/inhomogeneous 
medium. Increasing the gas-to-dust ratio, however, does not increase the 
predicted radio flux beyond $\sim5$ Jy. It is, therefore, probable that the 
difference is due non-uniform distributions of gas in this region.

\subsubsection{IRAS 14416-5937 - B}

By integrating the observed SED, the total luminosity obtained for IRAS 
14416-5937 - B is $6.8\times10^4$ L$_{\odot}$. The best fit 
radiative transfer model along with the observed SED is shown in Figure 
\ref{radtran} (right), and the parameters of the best fit model are presented 
in Table \ref{radtable}. The best fit model is a uniform density 
distribution of gas and dust. The outer size of the cloud is 2.8 pc and the 
optical depth at 100 $\mu$m is 0.018. We have used a single ZAMS star 
of spectral type O8-O7.5 to carry out the radiative transfer 
modelling. In the best fit model, the DL type of dust has been used. 

\subsection{Line emission from gas in IRAS 14416-5937 - A}

Since high resolution spectroscopic observations of a region near IRAS 
14416-5937 - A are available from ISO-LWS, an attempt has been made 
to model this source using a sophisticated scheme which includes the gas 
component with significant details. This scheme, based on CLOUDY, 
predicts infrared 
nebular/ionic fine structure line emission from the interstellar gas in IRAS 
14416-5937 - A, which has been compared with the ISO-LWS observations. 

In order to model line emission from gas, several prominent 
elements in the gas phase of the cloud have been considered. Physical 
processes like thermal balance considering various heating and cooling 
processes, photoionisation,
recombination, collisional excitation and de-excitation, grain photoionisation
and gas-dust coupling have been included in the model. The detailed modelling
involves the use of photoionisation code CLOUDY (Ferland, \cite{Fe96}) which 
has been supplemented with a software scheme developed by Mookerjea \& Ghosh 
(\cite{Mo99}).
This scheme improves the modelling by (a) emulating the exact structure of
 the HII region and (b) including absorption effects of dust (present within
the line emitting zones) on the emergent line intensities. Typical HII region
abundance of the gas component, tabulated by Ferland (\cite{Fe96}) has been taken
into consideration. Elements with relative abundance higher than
$3.0\times10^{-6}$ have been used; these are H, He, C, N, O, Ne, Mg, Si, S and
Ar.

The geometry of the cloud has been taken to be identical to that 
obtained from modelling of the continuum SED (see subsection 4.1.1). CLOUDY is 
run twice. First, it
is for the pure gas inner shell. The emerging spectrum comprises of continuum
and line emission. This emerging continuum from the inner shell is input to
the second shell comprising of gas and dust. The line emission from the inner
shell in the first run is transported outside through extinction by the dust
column in the second shell. The emerging line luminosities from both the shells
are finally added to predict the total luminosity. A total of 27 spectral
lines in the wavelegth range $2.5-200$ $\mu$m have been considered. The
predicted emerging spectrum has been computed by convolving the
spectral lines with typical spectral resolutions of ISO-SWS and LWS, viz.,
1000 for $2.5\le\lambda<12$ $\mu$m, 20000 for $12\le\lambda<45$ $\mu$m, 8100
for $45\le\lambda<75$ $\mu$m and 6800 for $75\le\lambda<200$ $\mu$m.

For IRAS 14416-5937 - A, the emergent spectrum (obtained using the above 
procedure) shows a total of 18 nebular/ionic lines satisfying our 
detectability criterion (power in the line $>$ 1 \% of power in the 
neighbouring continuua). The wavelengths and luminosities of these lines are 
presented in Table \ref{model_line}. The ratio of luminosity of each line to 
that of [C II] line at 158 $\mu$m is also listed. The complete emerging 
spectrum, including lines from the 10 elements considered as well as the 
continuum predicted by this model, is shown in Figure \ref{cloudy}. 

\section{Discussion}

\subsection{Emission from dust}

\begin {figure*}
\hskip -1cm
\includegraphics[height=7.0cm]{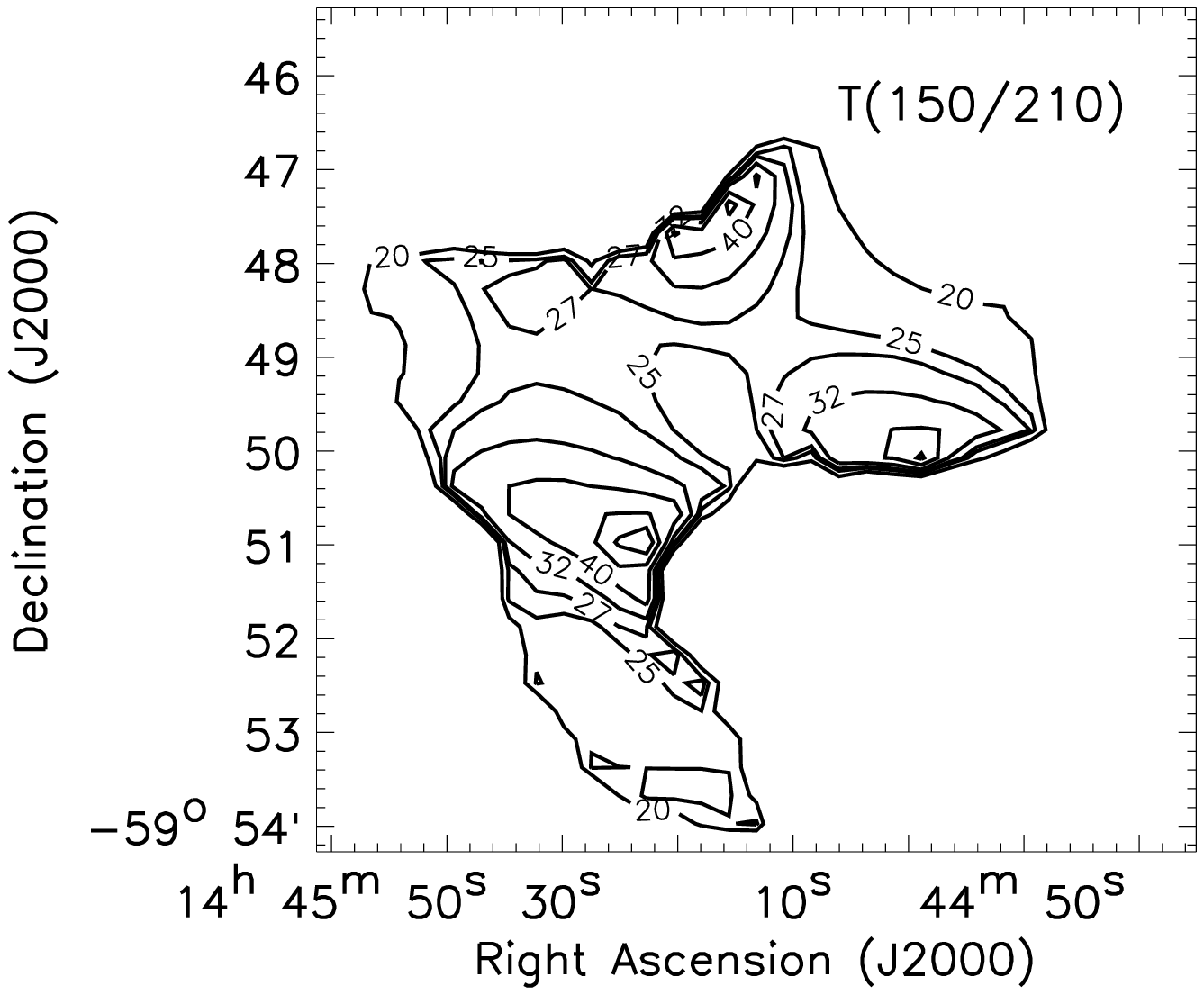}
\hspace*{-2cm}
\includegraphics[height=7.0cm]{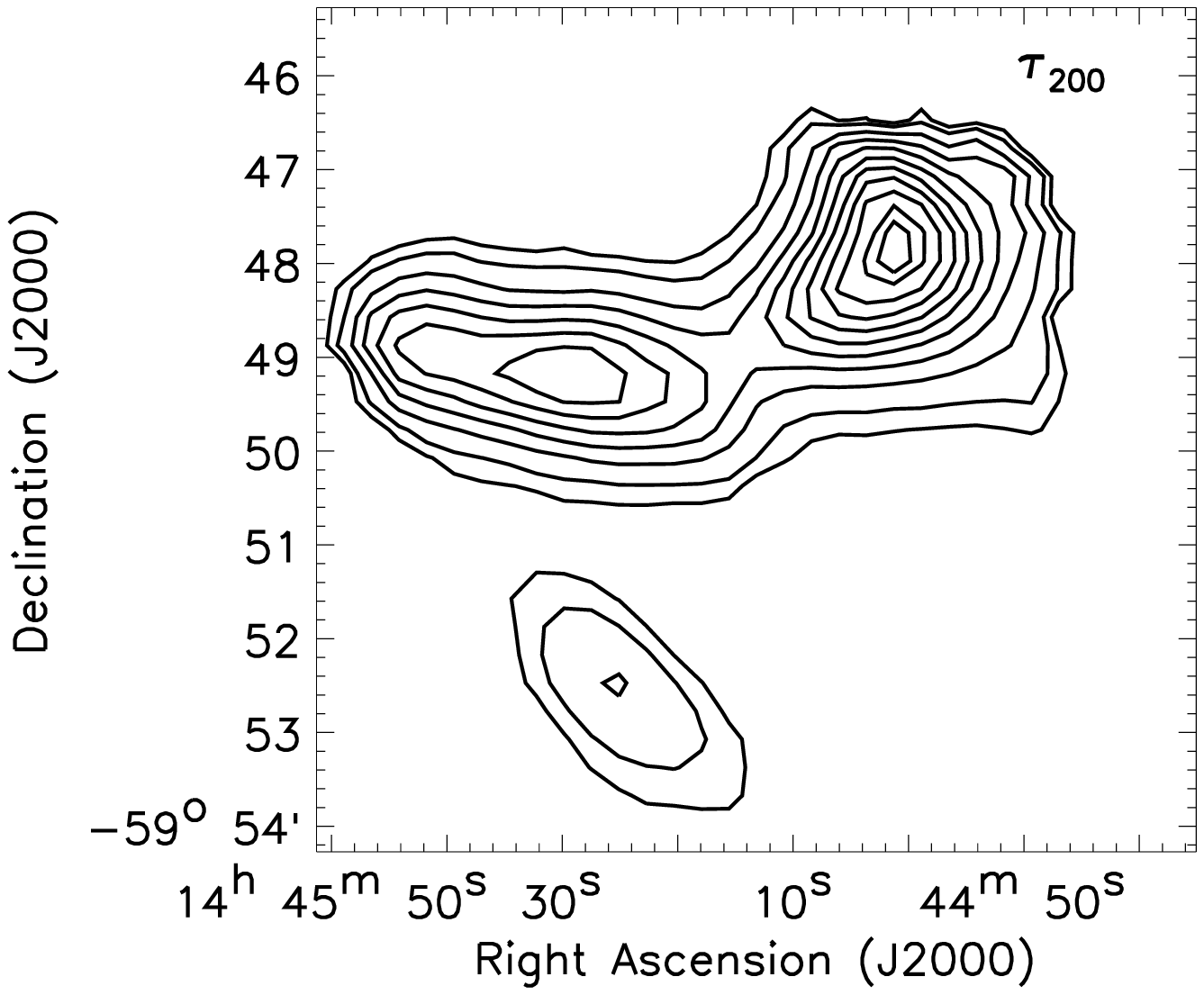}
\caption{The distribution of dust temperature T(150/210) (left), and
optical depth at 200 $\mu$m, $\tau_{200}$, (right) from the region around IRAS
14416-5937 assuming a dust emissivity law of $\epsilon_{\lambda} \propto
\lambda^{-2}$. The isotherms correspond to 20, 25, 27, 32, 40, 50 and 60 K. The
$\tau_{200}$ contours represent  5, 10,  20, 30, 40, 50, 60, 70, 80 and 90 \% of the peak value of $0.06$.
}
\label{Ttaubal}
\end {figure*}

\begin {figure*}
\hskip 3cm
\includegraphics[height=8.0cm]{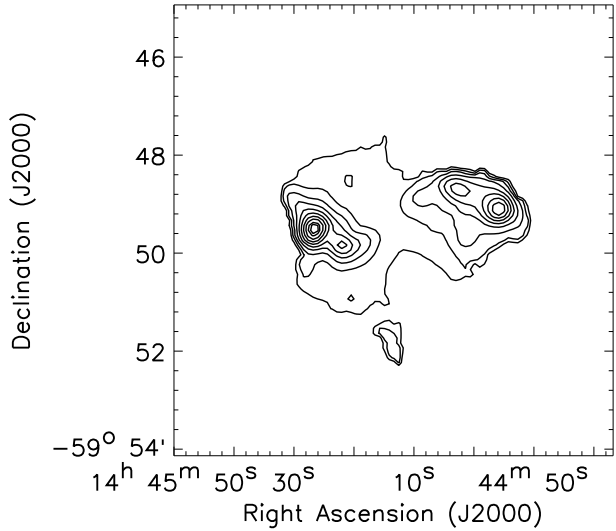}
\caption{The emission in UIBs for the region around IRAS 14416-5937. The 
contour levels are at 1, 5, 10, 20, 30, 40, 50, 60, 65, 70, 80, 90, 95\% of 
peak value of $1.3\times10^{-4}$ W m$^{-2}$ Sr$^{-1}$. 
}
\label{pah}
\end {figure*}

\begin {figure*}
\hskip 3cm
\includegraphics[height=8.0cm]{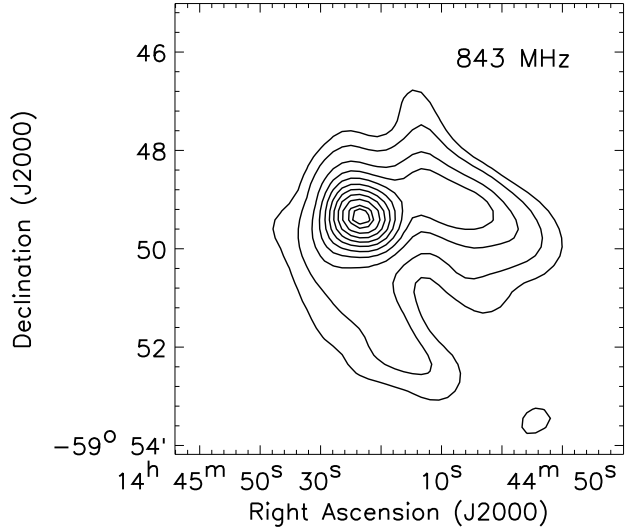}
\caption{The SUMSS radio flux density map for the region around IRAS 14416-5937
 at 843 MHz. The contour levels are at 5,
10, 20, 30, 40, 50, 60, 70, 80, 90, 95\% of peak flux of 4.4 Jy/beam. The
beam is $\sim43\arcsec\times50\arcsec$.
}
\label{SUMSS}
\end {figure*}

\begin{figure*}
\begin{center}
\includegraphics[height=8.0cm]{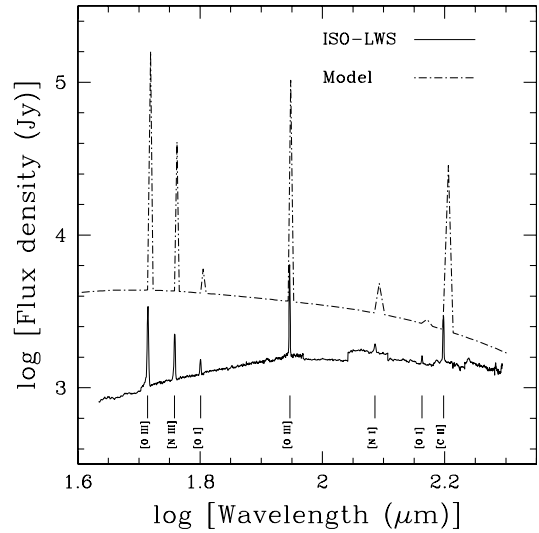}
\caption{ISO-LWS spectrum (solid line) shown alongwith the model calculations 
(dash-dotted line). The ISO-LWS spectrum is taken for a region $\sim1.4'$ to 
the north-west of IRAS 14416-5937 - A. However, the model calculations are 
carried out for IRAS 14416-5937 - A.
}
\label{ISO_cloudy}
\end{center}
\end{figure*}

\begin {figure*}
\includegraphics[height=7.5cm]{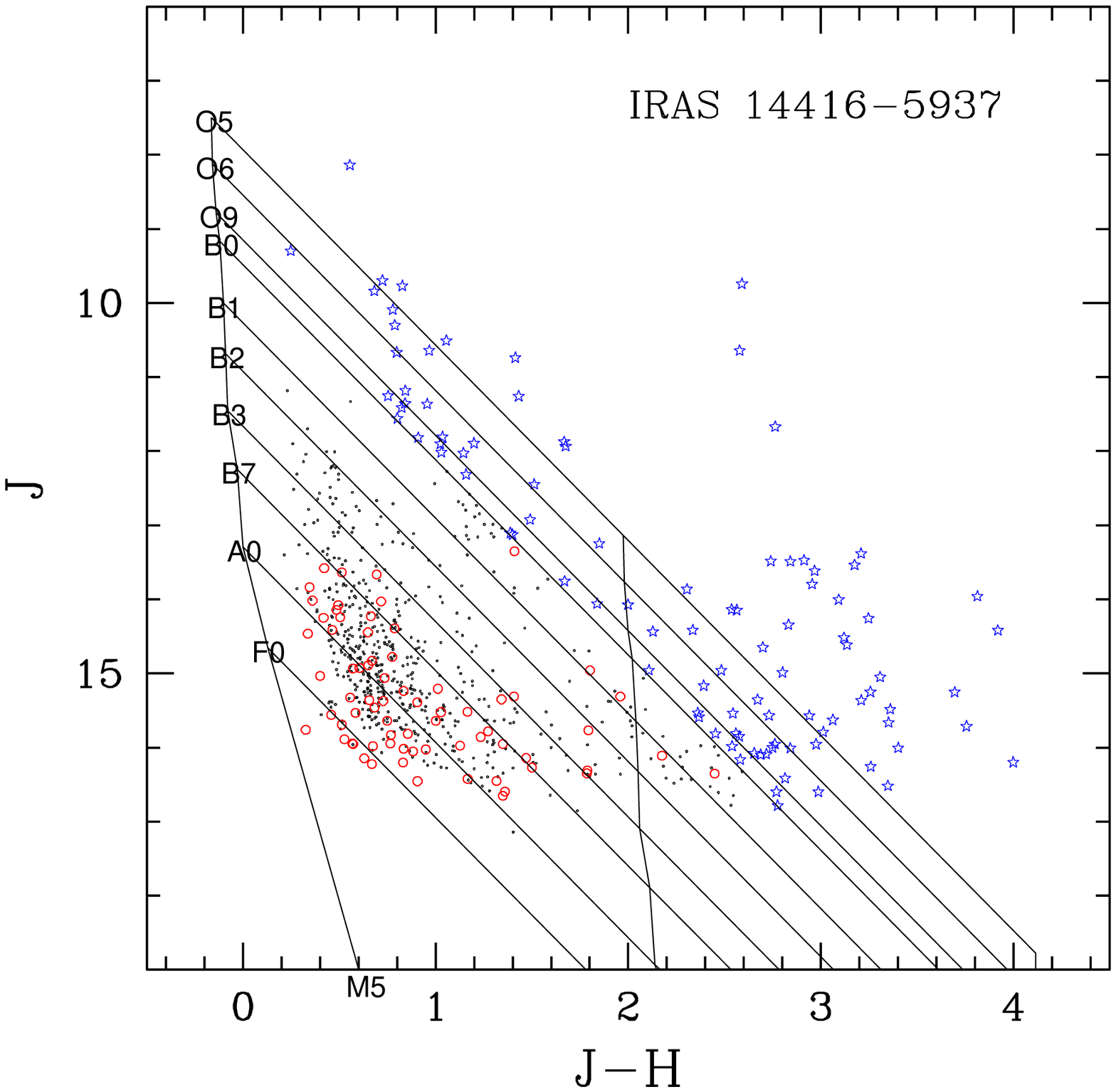}
\hskip -0.5cm
\includegraphics[height=7.5cm]{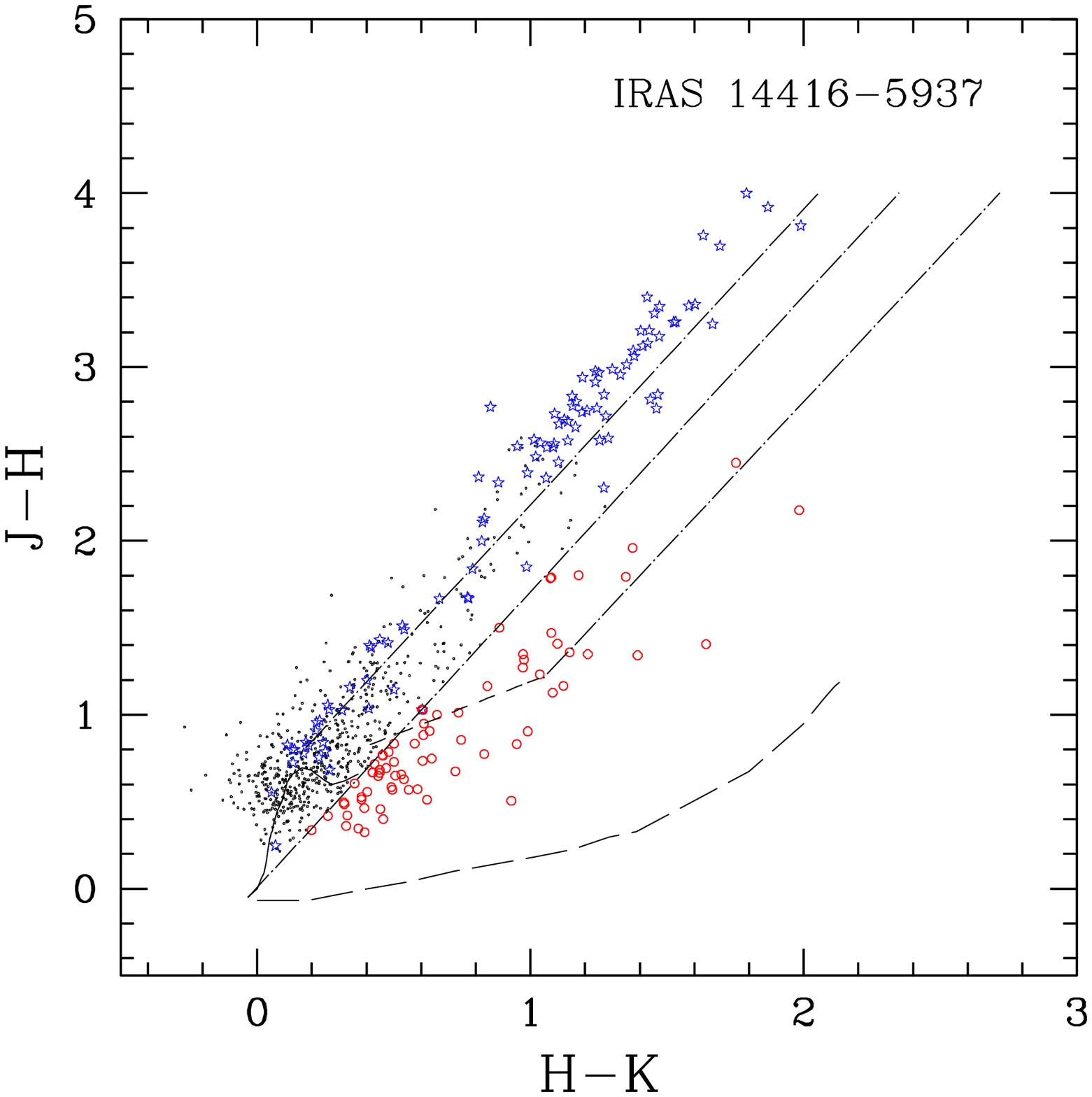}
\caption{ Colour-magnitude (left) and colour-colour diagram (right) for sources
 detected in all the three 2MASS bands for the region around IRAS 14416-5937. 
In the colour-magnitude diagram, the nearly vertical solid lines from left to 
right represent the zero age main sequence (ZAMS) curves reddened by A$_V$ = 0, 
and 20 magnitudes respectively. The slanting lines trace the reddening 
vectors of these ZAMS stars. In the colour-colour diagram, the locii of the 
main sequence and giant branches are shown by the solid and dotted lines 
respectively. The short-dash line represents the locus of T-Tauri stars. The 
three parallel dash-dotted straight lines follow the reddening 
vectors of giants, 
main sequence stars (or dwarfs) and the T-Tauri stars. The long 
dashed line represents the locus 
of Herbig Ae/Be stars. The asterisk symbols represent sources lying above the 
ZAMS curve B0. The open circles represent sources depicting an infrared excess in the colour-colour diagram. The dots are sources lying below the 
ZAMS spectral curve of B0 (see text for details).
}
\label{cmcc_2mass}
\end {figure*}

\begin {figure*}
\hskip 3cm
\includegraphics[height=8.0cm]{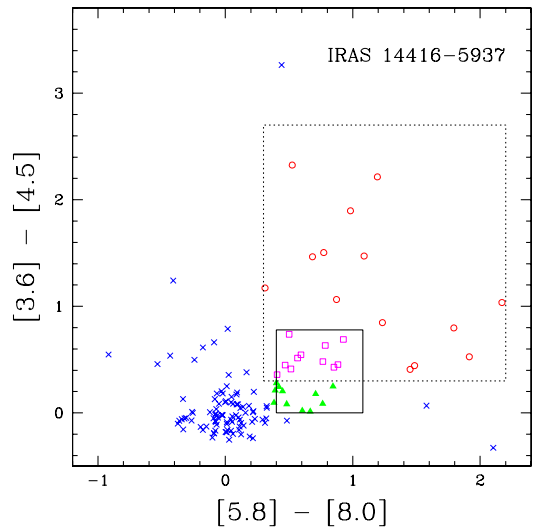}
\caption{Colour-colour diagram of the sources detected in all the four 
Spitzer-IRAC bands for the region around IRAS 14416-5937. The open circles, open squares, filled triangles and the cross symbols denote Class I, Class I/II, 
Class II and other sources, respectively. The solid-line square in the diagram 
approximately delineates the region occupied by the Class II sources
whereas the dotted-line square covers the region occupied by the Class I 
sources as shown in the models of Allen et al. (\cite{Al04}) 
}
\label{cc_glim}
\end {figure*}

\begin {figure*}
\includegraphics[height=7.5cm]{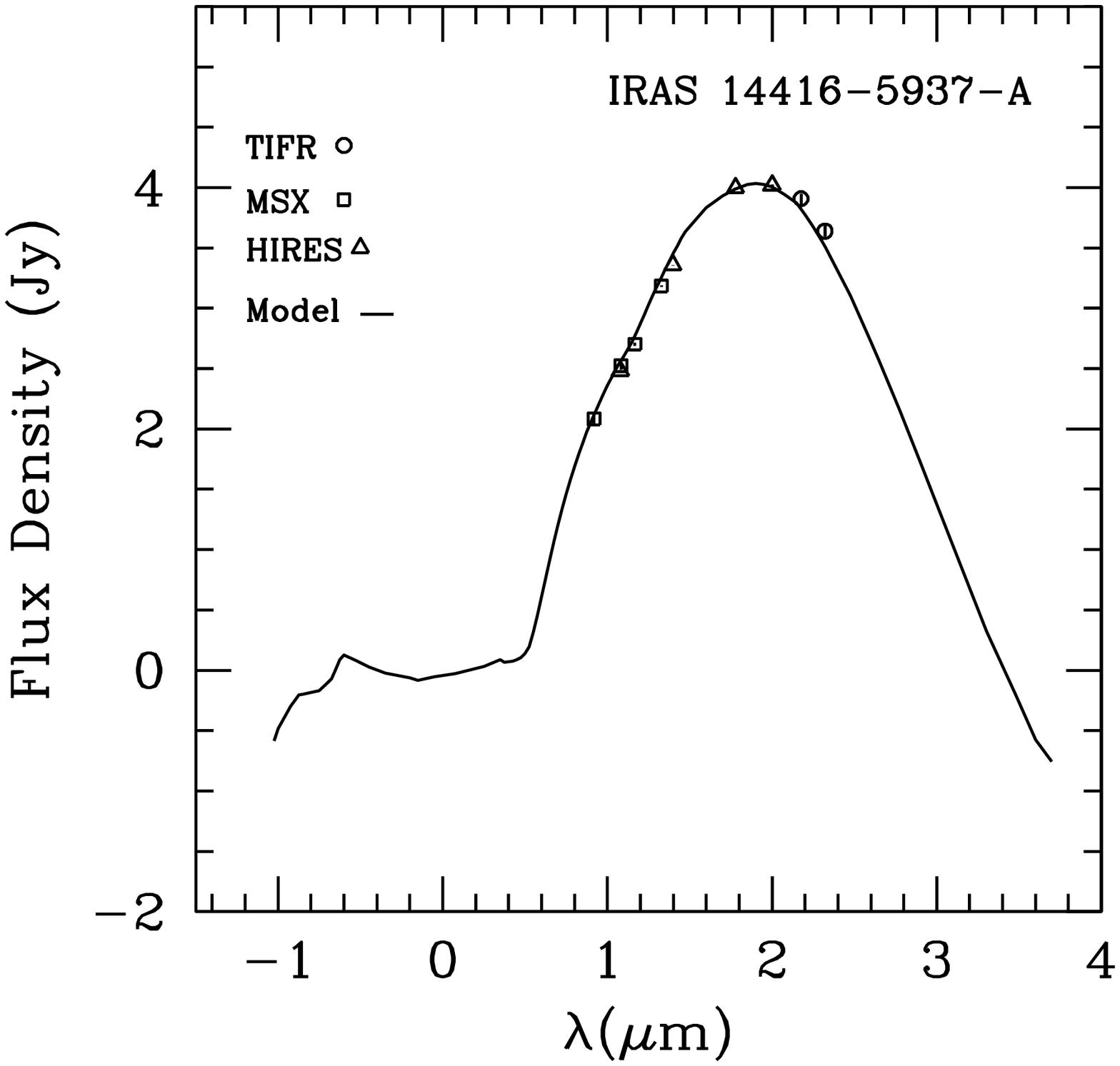}
\includegraphics[height=7.5cm]{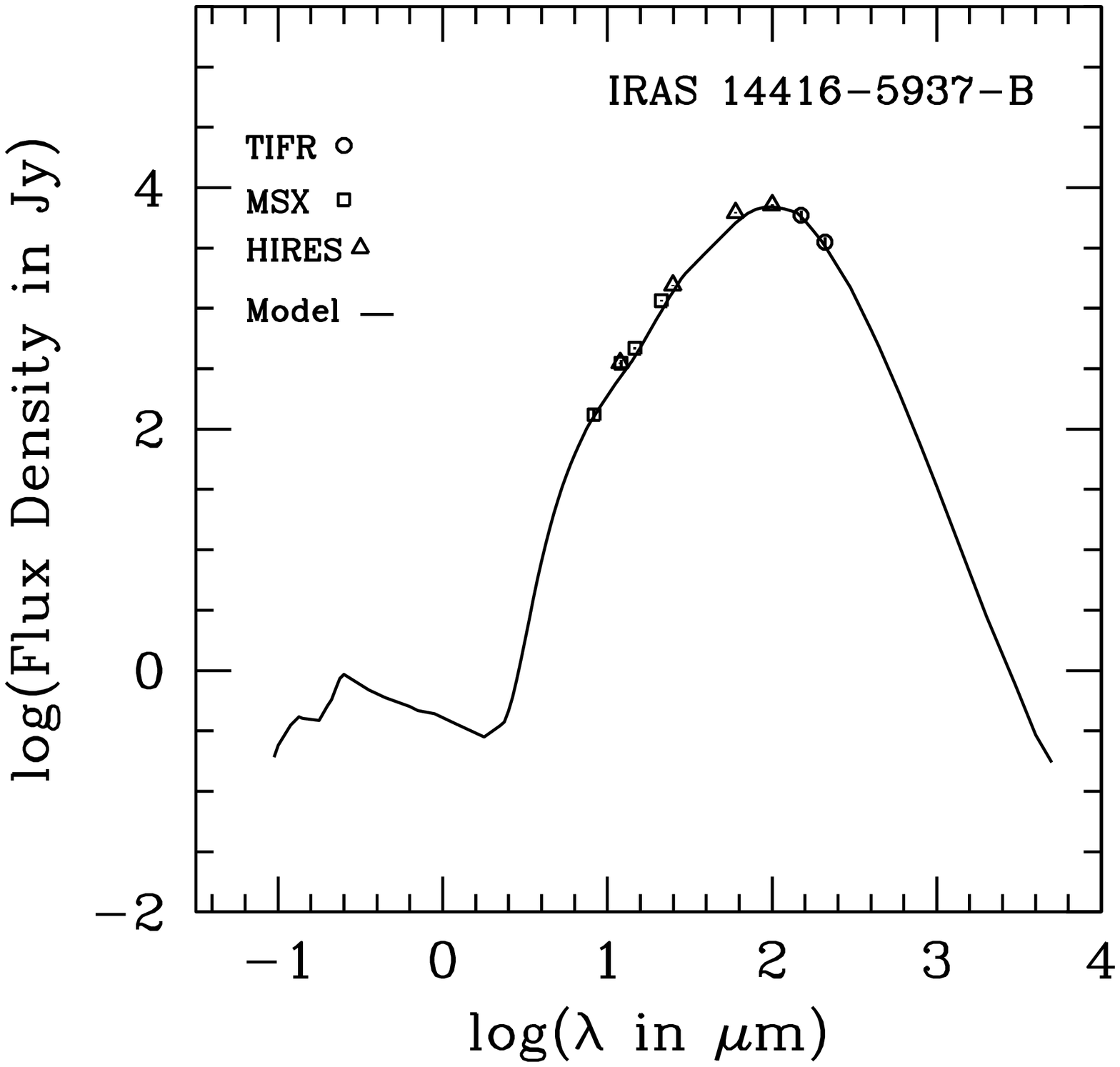}
\caption{Comparison of the spectral energy distribution from observations and 
the best fit radiative transfer model of IRAS 14416-5937 - A (left) and 
IRAS 14416-5937 - B (right). The open circles, triangles and squares 
represent the TIFR, IRAS-HIRES and MSX data respectively. The solid line 
is the best fit radiative transfer model to the data. See text and Table 
\ref{radtable} for details of the model parameters.
}
\label{radtran}
\end {figure*}

\begin{figure*}
\begin{center}
\includegraphics[height=8.0cm]{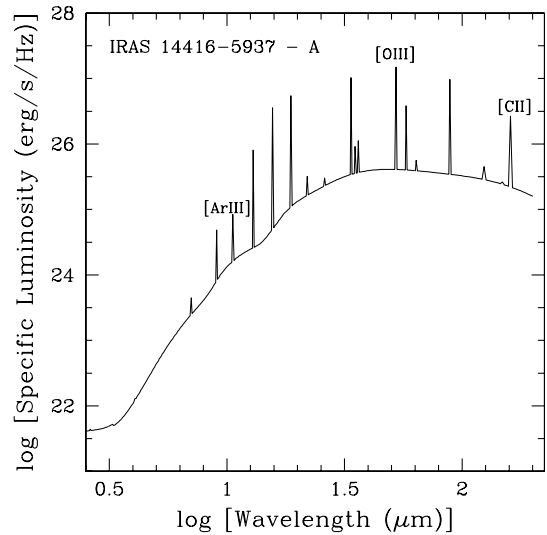}
\caption{Emergent spectrum from model calculations of IRAS 14416-5937 - A. 
Few lines are identified and the details of the lines are given Table 
\ref{model_line}. }
\label{cloudy}
\end{center}
\end{figure*}

The far infrared TIFR (150 \& 210 $\mu$m) and IRAS-HIRES (60 \& 100 $\mu$m) 
maps probe emission from relatively colder dust in the complex including  
IRAS 14416-5937 A \& B. The temperature of the far infrared emitting cold dust 
is $\sim 25-30$ K (Figure \ref{Ttaubal}). It is interesting to note that, of 
the IRAS-HIRES images in the four wavebands, the peak emission at 12 $\mu$m is at B while 
for the other three bands (25, 60 \& 100 $\mu$m), it is at A. 
From the IRAS-HIRES maps too, we see extended dust emission. Also, in all the 
six maps, there is extended diffuse emission towards the south of 
source A. The flux densities from the TIFR maps at 150 and 210 $\mu$m have 
been used to compute the mass of the dust component using the formulation of 
Hildebrand 
(\cite{Hi83}) and Sandell (\cite{Sa00}). For a temperature of 27 K and 24 K, 
obtained for IRAS 14416-5937 A and B (see Figure 
\ref{Ttaubal}), we find their dust masses to be 31 and 36 M$_{\odot}$, 
respectively. This is in fairly good agreement with the dust masses 
of 25 and 49 M$_{\odot}$ obtained from the radiative transfer modelling. 

From Figure \ref{Ttaubal}, we see that the observed peak optical depth at 200 
$\mu$m is 0.06 close to B. However, we find from radiative transfer modelling, 
that the optical depth at 100 $\mu$m is 0.02. This apparent 
inconsistency (of lower value of $\tau_{100}$ compared to $\tau_{200}$) 
could possibly indicate a clumpy/inhomogeneous medium and/or dust 
grains of different properties than used here.
The modelled UIB emission (Figure \ref{pah}) peaks close to A. The UIB 
emission map shows that both A and B are resolved into two sources each. This 
is due to higher angular resolution of the MSX maps compared to the 
IRAS-HIRES and TIFR maps. These secondary peaks could be due to locally higher 
radiation fields caused by early type stars (since the UIB is 
expected to be 
excited by UV photons). To locate the possible sources responsible for exciting
 the secondary UIB peaks, we looked at the sources in 2MASS and Spitzer 
catalogs. However, we do not find any such candidates around these peaks. This 
may perhaps be due to the high extinction around IRAS 14416-5937 
region.  

\subsection{Emission from gas}

We compare radio continuum emission from the SUMSS radio map 
at 843 MHz with the high angular resolution map at 6.67 GHz of Walsh et al. 
(\cite{Wa98}). The high angular resolution ($\sim1.5''$) map at 6.67 GHz shows 
two main peaks at ($\alpha_{2000},\delta_{2000}$) = ($14^h$ $45^m$ 
$22.54^s$, $-59^{\circ}$ 49$\arcmin$ 37.3$\arcsec$) and ($\alpha_{2000}, 
\delta_{2000}$) = ($14^h$ $45^m$ $22.25^s$, $-59^{\circ}$ 49$\arcmin$ 
39.5$\arcsec$). These are nearly $14''$ and $17''$ offset from the SUMSS peak 
at 
843 GHz, respectively. This could be either due to the effect of 
lower angular 
resolution of the SUMSS map or the effect of opacity. It is also important to 
note that the radio map at 6.67 GHz at high angular resolution covers only the 
very compact features in this HII region. A comparison of the radio 
peaks with the peaks of UIB emission shows that the 
secondary peak near A of the UIB emission is close ($\sim12''$) to the high 
angular resolution radio peaks at 6.67 GHz of Walsh et al. (\cite{Wa98}).

In the ISO-LWS spectrum, we notice that the cooling lines from the Photo-Dissociated Regions (PDRs): [O I] 63, 
145 $\mu$m and [C II] 158 $\mu$m have been clearly detected. In addition, lines coming 
from the higher excitation potential ions such as [N II] 122 $\mu$m, [N III] 57
 $\mu$m as well as [O III] 52 and 88 $\mu$m lines are also detected. The maximum flux is observed in the [O III] 52 $\mu$m line. Since the 
spectrum is taken at a position which is located around $\sim1.4'$ to the 
north-west of IRAS 14416-5937 - A, it is likely that the observed fluxes in 
the highly 
ionised species like [O III]  and [N III] are due to  an extended 
component of 
low density ionised gas rather than compact source(s). In the highest density 
regions of compact cores, these lines are collisionally de-excited (Morisset 
et al., \cite{Mo02}). The ratio of fluxes in [O III] lines has been 
used to estimate the electron density, $n_e\sim 300$ cm$^{-3}$ in this region 
(details are given in Appendix A). 
Using the fluxes in [N III] 57 $\mu$m and [N II] 122 $\mu$m lines, the 
effective temperature of the ionising radiation is found to be $\sim37,500$ K
(details in Appendix A). This compares well with the effective 
temperature, $39,500$ K, of the centrally exciting source required by the 
radiative transfer model (see Section 4.1.1). Next, we consider the flux ratio 
of [N III] 57 
$\mu$m to [O III] 52 $\mu$m. Since the ionisation potentials of these two 
ions are similar (O$^{++}$ = 35.1 eV; N$^{++}$ = 29.6 eV), they are likely 
to sample the same volume of gas (Mizutani et al., \cite{Mi02}). In our case, 
the intensity ratio from the observed ISO-LWS spectrum is 
I([N III] 57)/I([O III] 52) $\sim 0.40$.
Mizutani et al. (2002) find that this ratio is almost constant and 
obtain the value of $\sim0.3$ for the optically bright regions of the Carina 
nebula and a value ranging from $0.25-0.5$ for the surrounding region.  
We have also estimated the density and the radiation field using the ratio 
([C II] + [O I]$^{63}$)/$L_{FIR}$ which is a measure of the gas heating 
efficiency and the ratio of line intensities [C II]/[O I]$^{63}$. The 
total far infrared emission is obtained by integrating the ISO-LWS continuum. 
We obtain FIR flux for this region to be $7.1\times10^{-11}$ W m$^{-2}$.
 The gas density  
and radiation field are found to be $\sim100$ cm$^{-3}$ and $\sim300$ G$_0$ 
(G$_0$ is Habing Field = $1.6\times10^{-6}$ W m$^{-2}$) 
respectively from the Figure 4 (right) of Peeters et al., (\cite{Pe05}).
 
Figure \ref{ISO_cloudy} shows comparison of the spectra from ISO-LWS and 
the model calculations. The flux densities from the model 
calculations are higher than that observed from ISO. This is expected since the 
ISO-LWS beam is centered $1.4'$ away from IRAS 14416-5937 - A peak. Also, the 
ISO-LWS beam is   
of size $84''$ whereas the model computes the total emergent 
intensities from the entire cloud. It is instructive to compare the observed 
and modelled fine structure line ratios, normalised with respect to [C II] 
158 $\mu$m line. The ratio of the lines corresponding to the doubly ionised 
atoms [O III] 52 and 88 $\mu$m and [N III] 57 $\mu$m are overestimated by the 
model as compared to the observations by a factor of upto 2. On the other 
hand, the ratio of the lines [O I] 63 $\mu$m and [N II] 122 $\mu$m is 
underestimated by the model by a factor of $\sim3$. The major difference is found for the [O I] 145 $\mu$m line 
ratio which is underestimated by the model by a factor of $\sim16$. 

\subsection{Association with the cluster}

\begin {figure*}
\vspace*{-3cm}
\hskip 2cm
\includegraphics[height=15.0cm]{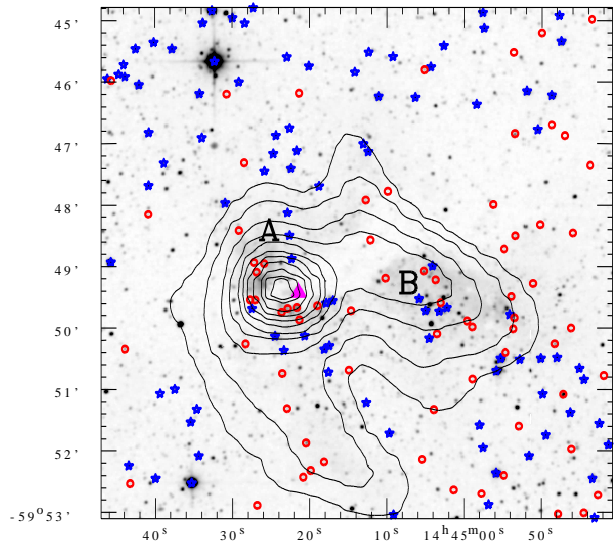}
\vspace*{-3cm}
\caption{The radio contours (SUMSS; 843 MHz) overlaid over the 2MASS 
K$_s$ band image for the region around IRAS 14416-5937. The labelled axes are 
in J2000 coordinates.  The regions `A' and `B' are shown and various near 
infrared sources are marked. The
 asterisk symbols represent sources of spectral type earlier than B0 and the 
 circles denote the infrared excess sources. The solid triangle represents J14452143-5949251 (2MASS source closest to radio peak detected in all the 
three (JHK$_s$) bands and of spectral type earlier than O6 from the CM diagram).}
\label{jhk_igi}
\end {figure*}

Next we consider the embedded star clusters associated with the IRAS 
14416-5937 region based on the near \& mid 
infrared emission. From the 2MASS CM diagram in Figure \ref{cmcc_2mass} (left),
we find that there are 98 sources lying above the reddening curve of 
the ZAMS spectral type B0 for a distance of 2.8 kpc. It is unlikely that all 
these objects are ZAMS stars of spectral type earlier than B0 and associated 
with the cluster since the combined luminosity of these stars would be
 much higher 
than the observed luminosity. It is reasonable to consider that many of these 
objects are foreground stars or bright background giants not associated with 
the star forming region although a few of these may be O-B stars belonging to 
the cluster. From the 2MASS CC diagram,
in Figure \ref{cmcc_2mass} (right), we find that there are 72 sources having 
an infrared excess (based on the CC diagram). These sources have 
been overplotted on the 
2MASS K$_s$ band image of the region around IRAS 14416-5937 in Figure 
\ref{jhk_igi}. The grayscale 2MASS image of this region shows diffuse 
emission apart from a number of sources. A dust lane $\sim3.5$ pc long running 
diagonally across 
the image commencing from the source A is clearly observed. As compared to the 
rest of the image where a number of 2MASS sources are detected, very few 
sources are seen in the region of the dust lane. From the spatial distribution 
of the sources in the Figure, we find that most of the 2MASS sources within 
the dust lane are those with infrared excess. A larger number of sources with 
infrared excess are present in the IRAS 14416-5937 - A region as compared to 
B. On the other hand, we find sources of `spectral type' earlier than 
B0 clustered near IRAS 14416-5937 - B. This suggests the possibility that the 
complex B could be more evolved than A. While the sources lying above the 
reddening vector of ZAMS spectral type B0 in the 2MASS CM diagram are 
distributed everywhere in the image other than the dust lane, we see a larger 
number of sources with infrared excess below the dust lane (i.e bottom right 
part of the image). These objects with infrared excess are likely 
to be pre-main 
sequence candidate objects in this star forming region. We have also compared 
the statistics of the infrared excess
 sources with that obtained from two neighbouring control fields of the same 
size. We find that the number of infrared excess sources in the star forming 
field is larger than those obtained from the control fields by a factor of 
$\sim2$. 

\begin {figure*}
\hskip 2cm
\includegraphics[height=6.5cm]{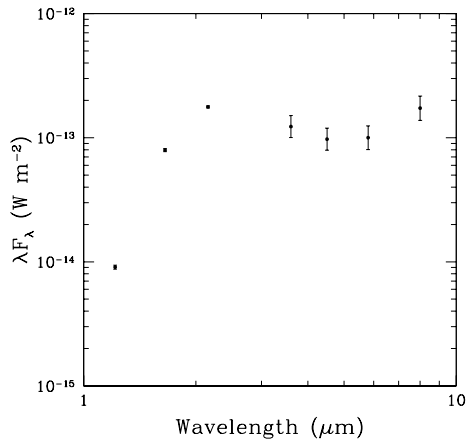}
\caption{The spectral energy distribution of J14452143-5949251 (2MASS source 
closest to radio peak detected in all the three (JHK$_s$) bands and of spectral type earlier than O6 from the CM diagram) constructed using the fluxes from 
2MASS and Spitzer-GLIMPSE (details in text).
}
\label{sed_radpk}
\end {figure*}

In Figure \ref{jhk_igi}, the SUMSS radio contours at 843 MHz have 
been overplotted on the grayscale K$_s$ band image of 2MASS. Within the radio 
nebulosity, we find that there are a number of asterisks (sources lying above 
the reddening vector of ZAMS spectral type B0 and without infrared excess in 
the 2MASS CM diagram). Within the 30\% contour level, there are six asterisks. 
The details of these sources are listed in Table \ref{O6_2mass}. Among these, 
J14452143-5949251 is closest to the radio peak ($\sim15.7''$). We, however, do 
not find a Spitzer-GLIMPSE counterpart of J14452143-5949251 within a search 
radius of $0.8''$ (search radius used for associating the 2MASS and 
Spitzer-GLIMPSE sources). An investigation of the IRAC images reveals strong 
emission at the 
location of this source in all the 4 IRAC (3.6, 4.5, 5.8 and 8.0 $\mu$m) bands. 
The spectral energy distribution of this source is shown in Figure 
\ref{sed_radpk}. The fluxes in the IRAC bands were obtained by 
integrating within an aperture of diameter 6$''$ (and applying the 
aperture corrections) centered on peak emission at 3.6 $\mu$m after 
subtracting the background. From the SED, we observe that this is a young 
reddened source with possibly a dust envelope/disk around it. The lower 
limit of the luminosity of this source obtained after integrating the SED is 
$\sim50$ L$_{\odot}$.
It would be interesting to do spectroscopy of this source, J14452143-5949251, 
in order to ascertain its spectral type and additional details.  

\begin {figure*}
\vspace*{-3cm}
\hskip 2cm
\includegraphics[height=15.0cm]{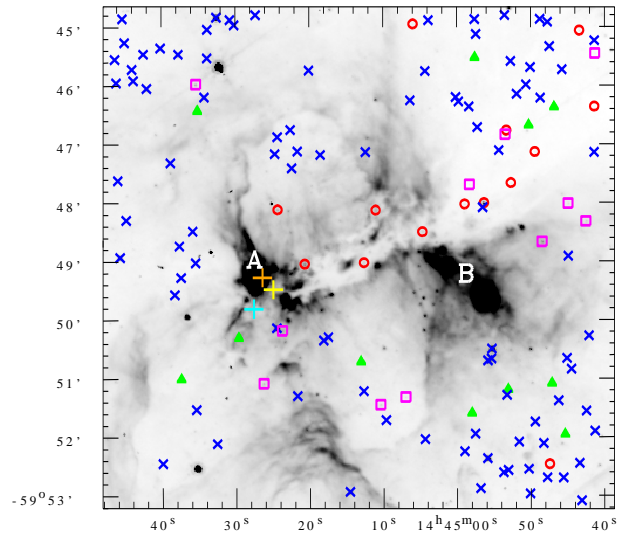}
\vspace*{-3cm}
\caption{ Spitzer-GLIMPSE sources detected in all the four IRAC bands 
overlaid over the Spitzer-GLIMPSE 5.8 $\mu$m band image for the region around 
IRAS 14416-5937. The labelled axes are in J2000 coordinates. The open 
circles, open squares, filled triangles and cross symbols represent the Class 
I, Class I/II, Class II and the other sources, respectively (see text for 
details).} The sources `A' and `B' are marked. The OH, H$_2$O and CH$_3$OH maser
 positions are shown by the light blue, yellow and orange plus-symbols, 
respectively.

\label{class_igi}
\end {figure*}

We have attempted to identify protostars and pre main-sequence objects among 
the Spitzer-GLIMPSE sources located around IRAS 14416-5937, detected in all 
the four IRAC bands, based on the models of Allen et al. (\cite{Al04}). These 
sources have been overplotted on the Spitzer-GLIMPSE 5.8 
$\mu$m band image, shown in Figure \ref{class_igi}. The Class I sources are 
shown as open circles, the Class I/II sources as open squares, Class II 
sources as filled triangles and the other sources as cross symbols. 
The details of these young Spitzer objects are available in Table \ref{elec1} 
(available only as electronic table).
We have also searched for 2MASS counterparts of these young GLIMPSE objects in 
this star forming region. A search radius of $0.8''$ has been taken as a 
criterion for associating the 2MASS and Spitzer-GLIMPSE objects.
While only 2 of the 14 Class I
 objects have 2MASS counterparts, 10 out of the 11 Class I/II objects have 
2MASS counterparts and all the 11 Class II objects have 2MASS 
counterparts. 
This is probably because most Class I objects suffer large amounts of 
extinction. We
also searched for Spitzer-GLIMPSE counterparts to the 2MASS infrared excess 
objects detected in all the bands of 2MASS (JHK$_s$). We find that 36 
out of  
72 2MASS IR excess objects have Spitzer-GLIMPSE counterparts. It is to be noted
that most of these objects are towards the lower right part of the image 
with few along the farther end of the dust lane. The details of these 2MASS IR 
excess objects with Spitzer GLIMPSE counterparts are given in Table \ref{elec2} 
(available only as electronic table).

The presence of diffuse near infrared emission around IRAS 14416-5937 - A and B in Figures \ref{jhk_igi} and \ref{class_igi} indicates 
that these are relatively evolved regions. 
The presence of HII regions around them is seen from the radio image
in Figure \ref{SUMSS}.
The OH (Caswell \& Haynes, \cite{Ca87}), H$_2$O (Caswell et al., \cite{Ca89}) 
and CH$_3$OH (Walsh et al., \cite{Wa98}) masers are marked in the figure as 
light blue, yellow and orange plus symbols, respectively. All these three 
masers are located close to A, near the peak of radio emission. The H$_2$O 
maser is closest ($\sim12''$) to the radio peak. The dust lane is very clearly 
seen in the 
IRAC images. It is interesting to note that the dust lane is lined up 
with only Class I sources. The dust lane has Class I sources extending upto 
$\sim5'$. It is to be further noted that Class I, Class I/II as well as 
Class II sources are present further along the dust lane and beyond B.
 Below B (i.e. the bottom right part of the image), we find Class I/II and 
Class II sources. As seen earlier, most of the 2MASS infrared 
excess sources with Spitzer-GLIMPSE counterparts are also located here. It is 
likely that the sources here represent an earlier generation of stars in the 
sequence of star formation. The other sources (cross symbols) are distributed 
over the entire image but away from the dust lane.

From this distribution, the following scenario emerges : it is evident that 
star formation activity is in progress along the dust-lane. Hence the deeply 
embedded Class I sources are seen along the dust lane. This 
suggests a possibility of star formation being triggered by the 
expanding HII region. A radio map at high angular resolution with diffuse 
features would help in confirming this. Further along and below the dust lane, 
the emergence of Class II sources indicates an advanced stage of star formation 
here. It is not clear what could have caused the star formation further along 
the dust lane. It would be very interesting to map this region in 
molecular lines 
which trace high density like CS, etc. Imaging in molecular near infrared line 
of H$_2$ would help in detecting shocked gas, if any, along the edges of the lane. 

\section{Summary}

The massive star forming region associated with IRAS 14416-5937 has 
been studied using the infrared (near, mid and far) wavebands. The dust
and gas environment as well as the stellar sources of this 
region have been probed using data from the TIFR balloon-borne telescope, MSX, 
IRAS-HIRES, Spitzer-GLIMPSE, 2MASS and ISO. The spatial distribution of far 
infrared emission from cold dust at 150 \& 210  $\mu$m has been obtained 
alongwith the maps of optical depth ($\tau_{200}$) and colour temperature, 
T(150/210). Using MSX data, the emission from warm dust and UIBs in 
this region has been studied. This region comprises of two sources: A (east) 
and B (west) as well as a dust lane due north-west of A. Using
2MASS PSC as well as the GLIMPSE catalogs, the near and mid infrared sources 
associated with this region have been studied using colour-magnitude 
and colour-colour diagrams. The atomic fine structure lines from the ISO-LWS 
spectrum of a region close to A have been used in estimating the 
electron density ($n_e\sim300$ cm$^{-3}$) as well as the effective temperature 
of the 
ionising radiation (37,500 K) in this region. Self-consistent radiative 
transfer modelling constrained by observations has been carried out 
through spherical gas-dust clouds for both the sources (A and B). A constant 
radial density distribution ($n(r) \propto r^0$) is preferred.
The geometric details of the gas-dust clouds, the dust composition and
optical depths, etc. have been obtained for the best fit models. We have also carried 
out modelling of line emission from source A using CLOUDY. The line ratios 
obtained from the model have been compared with the ISO-LWS spectrum of the 
region close to A. The Spitzer-GLIMPSE sources detected in all the 
IRAC bands (3.6, 4.5, 5.8, 8.0 $\mu$m) have been classified into Class I (14), 
Class I/II (11) and Class II (11) sources based on the models of Allen et al 
(\cite{Al04}). Their 
spatial distribution shows that Class I sources line up the dust lane. 
The Class II sources 
are found further along the dust lane indicating this to be a more evolved 
region. This suggests that the star formation is occuring along the  
dust lane possibly triggered by the shocks of the expanding HII 
regions of A and B.

\begin{acknowledgements}
We thank the anonymous referee for useful suggestions which improved 
the paper.
It is a pleasure to thank the members of the Infrared Astronomy Group at
TIFR for their support during laboratory tests and balloon flight campaigns.
All members of the Balloon Group and Control Instrumentation Group of the TIFR
Balloon Facility, Hyderabad, are thanked for their technical support
during the flight. We thank M. Walmsley and L. Testi for many useful 
suggestions regarding the improvement of this work. We thank A. Walsh 
for providing us the radio continuum map of IRAS 14416-5937 at 6.67 GHz.

\begin{itemize}

\item {We thank IPAC, Caltech, for providing us the HIRES-processed IRAS 
products.}

\item {This research made use of data
products from the Midcourse Space Experiment. Processing of the data was
funded by the Ballistic
Missile Defense Organization with additional support from NASA Office of Space
Science. This research has also made use of the NASA/ IPAC Infrared Science
Archive, which is operated by the Jet Propulsion Laboratory, Caltech, under
contract with the NASA.}

\item {This publication makes use of data products from the Two Micron All Sky
Survey, which is a joint project of the University of Massachusetts and the
Infrared Processing and Analysis Center/California Institute of Technology,
funded by the NASA and the NSF.}

\item {This work is based in part on observations made with the
\textit{Spitzer Space Telescope}, which is operated by the Jet Propulsion
Laboratory, under NASA contract 1407.}

\item {The MOST is operated by the University of Sydney and supported in parts
 by grants from the Australian Research Council.}

\item {Based on observations with ISO, an ESA project with instruments funded 
by ESA Member States (especially the PI countries: France, Germany, the 
Netherlands and the United Kingdom) and with the participation of ISAS and 
NASA.}

\end{itemize}

\end{acknowledgements}

\begin{table*}
\caption{Flux density details of IRAS 14416-5937}
\vspace{1cm}
\label{fluxes}
\begin{tabular}{|c |c| c c| c c c c |c c c c|} \hline \hline
 Source & Position & \multicolumn{10}{c|}{Flux Density$^a$ (Jy) for $\lambda$ ($\mu$m)} \\ \hline
 & (J2000) & \multicolumn{2}{c|}{TIFR images} & 
\multicolumn{4}{c|}{IRAS-HIRES images} & \multicolumn{4}{c|}{MSX images} \\ \hline
& & 210 & 150 & 100  & 60 & 25 & 12 & 21.3 & 14.7 & 12.1 & 8.3 \\ \hline
14416-5937 - A & 14$^h$ 45$^m$ 25.4$^s$ -59$^{\circ}$ 49$'$ 11$''$ & 4076 & 7545 & 9442 & 9345 & 2319 & 295 & 1528 & 504 & 334 & 121 \\
14416-5937 - B & 14$^h$ 45$^m$ 07.8$^s$ -59$^{\circ}$ 49$'$ 00$''$ & 3451 & 5601 & 6791 & 5750 & 1407 & 321 & 1159 & 468 & 351 & 132 \\ \hline
 \multicolumn{12}{|c|}{IRAS PSC}\\ \hline
14416-5937 & 14$^h$ 45$^m$ 27.9$^s$ -59$^{\circ}$ 49$'$ 14$''$ &   &      & 16100 & 6840 & 766 & 140 &  &  &  &  \\ \hline
\end{tabular}

$^a$ Fluxes obtained by integrating over a circular region of diameter $3\arcmin$. \\
\end{table*}

\begin{table}
\begin{center}
\caption{Fluxes derived from Gaussian fits to the fine-structure lines observed with ISO-LWS grating positioned at $14^h$ $45^m$
$21.0^s$   -59$^{\circ}$ 48$'$ 14$''$ (J2000).}
\label{iso_tab1}
\begin{tabular}{c c c c} \hline \hline
Element and & Wavelength & Flux & $\frac{f(line)}{f([C II])}$\\
Ionisation Stage & ($\mu$m) & ($10^{-17}$ W cm$^{-2}$) & \\ \hline
$[$C II$]$ & 157.78 & 1.23 & 1\\
$[$O I$]$  & 145.63 & 0.18  & 0.15\\
$[$N II$]$ & 121.84 & 0.36 & 0.29\\
$[$O III$]$ & 88.42 & 5.77 & 4.69\\
$[$O I$]$ & 63.23 & 0.68 & 0.55\\
$[$N III$]$ & 57.26 & 3.23 & 2.63\\
$[$O III$]$ & 51.85 & 8.29 & 6.74\\ \hline
\end{tabular}
\end{center}
\end{table}

\begin{table*}
\begin{center}
\caption{Best-fit parameters of the radiative transfer models for IRAS
14416-5937 - A and 14416-5937 - B.}
\vspace*{1cm}
\hspace*{-1.5cm}
\label{radtable}
\begin{tabular}{|c c c c c c c c c|} \hline \hline
Source & $\alpha$ & $R_{max}$ & $R_{min}$ & $r_{H II}$ & $\tau_{100}$ &
L & Dust & M$_{dust}$ \\
IRAS   &          &   (pc)    &   (pc)    &   (pc)    &              &
(10$^5$ $L_\odot$)  &  Type &  ($M_\odot$) \\ \hline
14416-5937 - A & 0.0  & 3.2 & 0.025 & 0.5 & 0.007 &  1.3 & DL & 25 \\
14416-5937 - B & 0.0  & 2.8 & 0.007 & 0.2 & 0.018 & 0.7 & DL & 49 \\ \hline
\end{tabular}
\end{center}
\end{table*}

\begin{table*}
\begin{center}
\caption{Emergent line luminosities predicted by the model for IRAS
14416-5937 - A}
\label{model_line}
\begin{tabular}{c r r r} \hline \hline
Element and & Wavelength & Luminosity & $\frac{L(line)}{L([C II])}$\\
Ionisation Stage & ($\mu$m) & ($L_{\odot}$) & \\ \hline
$[$C II$]$ & 157.78 & 17.91 & 1\\
$[$O I$]$  & 145.63 & 0.17  & 0.01\\
$[$N II$]$ & 121.84 & 1.57 & 0.09\\
$[$O III$]$ & 88.42 & 121.40 & 6.78\\
$[$O I$]$ & 63.23 & 2.64 & 0.15\\
$[$N III$]$ & 57.26 & 57.70 & 3.22\\
$[$O III$]$ & 51.85 & 271.54 & 15.16\\
$[$Ne III$]$ & 36.04 & 8.20 & 0.46\\
$[$Si II$]$ & 34.84 & 6.23 & 0.35\\
$[$S III$]$ & 33.50 & 116.71 & 6.52\\
$[$O IV$]$ & 25.91 & 1.12 & 0.06\\
$[$Ar III$]$ & 21.84 & 2.79 & 0.16\\
$[$S III$]$ & 18.69 & 111.75 & 6.24\\
$[$Ne III$]$ & 15.57 & 88.77 & 4.96\\
$[$Ne II$]$ & 12.82 & 23.81 & 1.33\\
$[$S IV$]$ & 10.52 & 50.65 & 2.83\\
$[$Ar III$]$ & 8.99 & 35.51 & 1.98\\
$[$Ar II$]$ & 7.00 & 2.23 & 0.12\\ \hline \hline
\end{tabular}
\end{center}
\end{table*}

\begin{table*}
\begin{center}
\caption{Details of 2MASS PSC sources lying above the reddening vector of 
the ZAMS spectral type O6 in the 2MASS (J-H vs. J) CM diagram and lying within 
the 30\% contour of the radio peak
 at 843 MHz (see Figure \ref{jhk_igi} and text for details).}
\label{O6_2mass}
\vspace*{0.2cm}
\hskip 2cm 
\begin{tabular}{|c c c c c c|} \hline \hline
2MASS PSC & $\alpha_{2000}$ & $\delta_{2000}$ & J & H & K$_s$\\
designation  & (deg) & (deg) & (mag) & (mag) & (mag) \\ \hline
J14450407-5948596 & 221.266960 & -59.816566 & $11.17\pm0.02$ & $9.80\pm0.02$ 
& $9.33\pm0.03$ \\
J14450584-5949317 & 221.274341 & -59.825478 & $11.83\pm0.02$ & $10.22\pm0.02$  & $9.44\pm0.02$ \\
J14451702-5949336 & 221.320956 & -59.826023 & $13.40\pm0.03$ & $10.31\pm0.03$ &
 $8.84\pm0.03$ \\
J14451788-5949360 & 221.324514 & -59.826691 & $15.88\pm0.19$ & $13.11\pm0.08$ &  $11.66\pm0.08$ \\
J14452143-5949251 & 221.339322 & -59.823666 & $14.12\pm0.04$ & $10.95\pm0.03$ &  $9.30\pm0.02$ \\
J14452450-5950084 & 221.352090 & -59.835682 & $14.26\pm0.04$ & $10.43\pm0.02$ &  $8.58\pm0.02$ \\
\hline
\end{tabular}
\end{center}
\end{table*}

\begin{landscape}
\pagestyle{empty}
\begin{center}
\hspace* {-2cm}
{\footnotesize
\topcaption{\label{elec1}Young stellar objects (Class I, Class I/II, Class II)
from the Spitzer-GLIMPSE survey as identified in the IRAC colour-colour diagram
(see details in the text).
}
\tablefirsthead{%
\hline
Spitzer-GLIMPSE & $\alpha_{2000}$ & $\delta_{2000}$ & 3.6 $\mu$m & 4.5 $\mu$m & 5.8 $\mu$m & 8.0 $\mu$m & 2MASS PSC & J & H & K$_s$  \\ \hline}
\tablehead{%
\multicolumn{11}{l}{\small \sl continued from previous page} \\
\hline
Spitzer-GLIMPSE & $\alpha_{2000}$ & $\delta_{2000}$ & 3.6 $\mu$m & 4.5 $\mu$m & 5.8 $\mu$m & 8.0 $\mu$m & 2MASS PSC & J & H & K$_s$  \\ \hline}
\tabletail{%
\hline
\multicolumn{11}{r}{\small \sl continued on next page} \\
%\hline}
}
\tablelasttail{\hline}
\begin{supertabular}{c c c c c c c c c c c}
designation  & (deg) & (deg) & (mag) & (mag) & (mag) & (mag) & designation & (mag) & (mag) & (mag) \\ \hline \multicolumn{11}{c}{Class I} \\ \hline
 G316.7459+00.0267 & 221.172131 & -59.772540 & $13.58\pm0.07$ & $12.41\pm0.10$ & $12.04\pm0.14$ & $11.73\pm0.33$ & - & - & - & - \\
 G316.7589+00.0445 & 221.180593 & -59.750896 & $13.27\pm0.08$ & $11.77\pm0.10$ & $10.86\pm0.09$ & $10.09\pm0.08$ & - & - & - & - \\
 G316.7144-00.0711 & 221.197417 & -59.874477 & $13.49\pm0.11$ & $12.65\pm0.11$ & $11.42\pm0.21$ & $10.19\pm0.18$ & - & - & - & - \\
 G316.7558+00.0078 & 221.205784 & -59.785483 & $13.50\pm0.09$ & $12.03\pm0.09$ & $11.45\pm0.09$ & $10.77\pm0.09$ & - & - & - & - \\
 G316.7582-00.0031 & 221.219329 & -59.794321 & $14.55\pm0.14$ & $12.66\pm0.10$ & $11.22\pm0.12$ & $10.24\pm0.19$ & - & - & - & - \\
 G316.7657+00.0099 & 221.221919 & -59.779357 & $12.67\pm0.09$ & $11.20\pm0.12$ & $10.71\pm0.07$ & $9.63\pm0.08$ & J14445329-5946458 & - & - & $14.14\pm0.11$ \\
 G316.7628-00.0115 & 221.234588 & -59.800018 & $10.62\pm0.08$ & $8.40\pm0.11$ & $7.86\pm0.05$ & $6.67\pm0.04$ & - & - & - & - \\
 G316.7676-00.0142 & 221.245584 & -59.800395 & $14.22\pm0.12$ & $11.89\pm0.08$ & $10.85\pm0.11$ & $10.33\pm0.24$ & - & - & - & - \\
 G316.7752-00.0264 & 221.269486 & -59.808246 & $14.13\pm0.21$ & $13.33\pm0.26$ & $10.89\pm0.12$ & $9.09\pm0.15$ & - & - & - & - \\
  G316.8026+00.0262 & 221.274722 & -59.749026 & $12.11\pm0.10$ & $11.67\pm0.11$ & $11.35\pm0.15$ & $9.86\pm0.18$ & J14450598-5944563 & - & $13.40\pm0.04$ & $12.81\pm0.03$ \\
  G316.7899-00.0264 & 221.295964 & -59.802072 & $13.31\pm0.15$ & $12.90\pm0.22$ & $10.19\pm0.15$ & $8.74\pm0.13$ & - & - & - & - \\
  G316.7865-00.0414 & 221.302607 & -59.817058 & $14.68\pm0.28$ & $13.64\pm0.24$ & $10.83\pm0.15$ & $8.66\pm0.17$ & - & - & - & - \\
  G316.8152-00.0381 & 221.351478 & -59.801898 & $12.02\pm0.07$ & $10.95\pm0.10$ & $9.99\pm0.09$ & $9.11\pm0.13$ & - & - & - & - \\
  G316.8017-00.0489 & 221.336176 & -59.817405 & $12.30\pm0.13$ & $11.77\pm0.21$ & $9.10\pm0.13$ & $7.18\pm0.23$ & - & - & - & - \\ 
\hline
\multicolumn{11}{c}{Class I/II} \\ \hline
 G316.7521+00.0405 & 221.171797 & -59.757430 & $12.37\pm0.07$ & $11.82\pm0.09$ & $11.67\pm0.13$ & $11.08\pm0.14$ & J14444123-5945268 & $15.53\pm0.05$ & $14.24\pm0.04$ & $13.45\pm0.04$ \\
 G316.7343-00.0040 & 221.176931 & -59.805271 & $11.34\pm0.05$ & $10.98\pm0.07$ & $10.46\pm0.11$ & $10.05\pm0.27$ & J14444244-5948188 & - & - & $13.44\pm0.04$ \\
 G316.7410-00.0015 & 221.187017 & -59.800188 & $11.51\pm0.06$ & $11.06\pm0.08$ & $10.70\pm0.09$ & $10.23\pm0.21$ & J14444488-5948006 & - & - & $13.96\pm0.05$ \\
 G316.7431-00.0145 & 221.201669 & -59.811088 & $11.10\pm0.07$ & $10.36\pm0.06$ & $9.60\pm0.07$ & $9.10\pm0.27$ & J14444839-5948400 & - & $15.58\pm0.16$ & $13.34\pm0.06$ \\
 G316.7655+00.0087 & 221.222566 & -59.780561 & $10.27\pm0.08$ & $9.84\pm0.07$ & $9.63\pm0.05$ & $8.78\pm0.05$ & J14445341-5946501 & $15.20\pm0.05$ & $13.31\pm0.04$ & $11.94\pm0.04$ \\
 G316.7687-00.0085 & 221.242800 & -59.794799 & $12.79\pm0.06$ & $12.10\pm0.10$ & $11.68\pm0.13$ & $10.75\pm0.20$ & J14445828-5947413 & - & - & $14.44\pm0.09$ \\
 G316.7596-00.0711 & 221.279027 & -59.855398 & $12.16\pm0.08$ & $11.65\pm0.11$ & $11.18\pm0.12$ & $10.61\pm0.07$ & J14450697-5951195 & - & $15.11\pm0.11$ & $13.84\pm0.05$ \\
 G316.7651-00.0761 & 221.293297 & -59.857536 & $11.06\pm0.08$ & $10.65\pm0.07$ & $10.09\pm0.08$ & $9.58\pm0.10$ & J14451038-5951271 & $16.03\pm0.11$ & $14.15\pm0.05$ & $13.03\pm0.03$ \\
 G316.7994-00.0688 & 221.348887 & -59.836427 & $12.04\pm0.12$ & $11.58\pm0.12$ & $10.39\pm0.10$ & $9.51\pm0.12$ & J14452376-5950111 & - & $14.81\pm0.12$ & $13.47\pm0.12$ \\
 G316.7978-00.0847 & 221.359476 & -59.851436 & $11.66\pm0.07$ & $11.18\pm0.10$ & $10.13\pm0.14$ & $9.36\pm0.30$ & - & - & - & - \\
 G316.8514-00.0155 & 221.397539 & -59.766099 & $11.08\pm0.05$ & $10.45\pm0.07$ & $9.43\pm0.07$ & $8.65\pm0.06$ & J14453543-5945581 & - & - & $14.05\pm0.07$ \\
\hline
\multicolumn{11}{c}{Class II} \\ \hline
 G316.7141-00.0614 & 221.188697 & -59.865808 & $11.30\pm0.07$ & $11.09\pm0.08$ & $10.94\pm0.09$ & $10.49\pm0.17$ & J14444533-5951567 & - & $13.63\pm0.10$ & $12.46\pm0.06$ \\
 G316.7563+00.0219 & 221.194869 & -59.772541 & $12.16\pm0.09$ & $11.87\pm0.13$ & $11.55\pm0.20$ & $11.15\pm0.28$ & J14444677-5946212 & - & - & $13.95\pm0.07$ \\
 G316.7236-00.0498 & 221.196152 & -59.851299 & $9.49\pm0.17$ & $9.28\pm0.07$ & $8.75\pm0.05$ & $8.36\pm0.09$ & J14444710-5951048 & $13.25\pm0.03$ & $11.90\pm0.04$ & $10.80\pm0.03$ \\
  G316.7607+00.0141 & 221.209268 & -59.777689 & $12.71\pm0.07$ & $12.62\pm0.10$ & $12.36\pm0.22$ & $11.60\pm0.25$ & J14445021-5946397 & - & $14.99\pm0.16$ & $13.58\pm0.07$ \\
 G316.7341-00.0566 &  221.220964 & -59.853038 & $11.43\pm0.04$ & $11.34\pm0.06$ & $11.22\pm0.11$ & $10.83\pm0.26$ & J14445303-5951111 & - & $13.92\pm0.06$ & $12.45\pm0.02$ \\
  G316.7827+00.0251 & 221.239761 & -59.758412 & $11.29\pm0.07$ & $11.20\pm0.08$ & $11.16\pm0.13$ & $10.68\pm0.22$ & J14445756-5945306 & - & $13.35\pm0.05$ & $12.31\pm0.04$ \\
 G316.7406-00.0671 & 221.241506 & -59.859770 & $10.38\pm0.05$ & $10.36\pm0.04$ & $10.14\pm0.08$ & $9.54\pm0.08$ & J14445797-5951352 & $15.44\pm0.06$ & $12.58\pm0.03$ & $11.39\pm0.03$ \\
  G316.7755-00.0672 & 221.304392 & -59.845108 & $11.04\pm0.04$ & $10.79\pm0.07$ & $10.43\pm0.08$ & $10.01\pm0.28$ & J14451304-5950424 & - & $14.28\pm0.06$ & $12.60\pm0.03$ \\
  G316.8098-00.0758 & 221.373630 & -59.838283 & $9.43\pm0.09$ & $9.19\pm0.08$ & $8.61\pm0.13$ & $7.77\pm0.34$ & J14452969-5950179 & - & $13.42\pm0.04$ & $11.10\pm0.02$ \\
  G316.8479-00.0221 & 221.396707 & -59.773534 & $11.47\pm0.08$ & $11.29\pm0.08$ & $11.03\pm0.10$ & $10.32\pm0.16$  & J14453522-5946248 & - & $14.62\pm0.07$ & $12.83\pm0.03$ \\
  G316.8197-00.0934 & 221.406214 & -59.850020 & $11.13\pm0.05$ & $11.12\pm0.06$ & $10.80\pm0.10$ & $10.13\pm0.13$  & J14453749-5951001 & $15.42\pm0.06$ & $12.95\pm0.03$ & $12.00\pm0.02$ \\
\hline
\end{supertabular}
}
\end{center}
%\end{table}
\end{landscape}

\begin{landscape}
\pagestyle{empty}
%\begin{table}
\begin{center}
{\footnotesize
\hspace* {-2cm}
\topcaption{\label{elec2}Infrared excess (young stellar objects) as identified
from the 2MASS colour-colour diagram which are also detected by Spitzer-GLIMPSE
(details in text).
}
\tablefirsthead{%
\hline
2MASS PSC & $\alpha_{2000}$ & $\delta_{2000}$ & J & H & K$_s$ & Spitzer-GLIMPSE & 3.6 $\mu$m & 4.5 $\mu$m & 5.8 $\mu$m & 8.0 $\mu$m \\
\hline}
\tablehead{%
\multicolumn{11}{l}{\small \sl continued from previous page} \\
\hline
2MASS PSC & $\alpha_{2000}$ & $\delta_{2000}$ & J & H & K$_s$ & Spitzer-GLIMPSE & 3.6 $\mu$m & 4.5 $\mu$m & 5.8 $\mu$m & 8.0 $\mu$m \\
\hline}
\tabletail{%
\hline
\multicolumn{11}{r}{\small \sl continued on next page} \\
%\hline}
}
\tablelasttail{\hline}
\begin{supertabular}{c c c c c c c c c c c}
designation  & (deg) & (deg) & (mag) & (mag) & (mag) & designation & (mag) & (mag) & (mag) & (mag)\\ \hline
J14444183-5950466 & 221.174299 & -59.846287 & $15.97\pm0.09$ & $15.14\pm0.11$ &  $14.52\pm0.11$ &   G316.7157-00.0406 & $14.15\pm0.14$ & $14.09\pm0.33$ & - & - \\ 
J14444256-5952434 & 221.177344 & -59.878723 & $15.76\pm0.08$ & $15.04\pm0.10$ &  $14.56\pm0.11$ &   G316.7035-00.0706 & $13.78\pm0.16$ & $14.23\pm0.19$ & - & - \\
J14444368-5947208 & 221.182020 & -59.789116 & $15.87\pm0.08$ & $15.36\pm0.11$ &  $14.84\pm0.13$ &   G316.7434+00.0096 & $14.58\pm0.16$ & $14.47\pm0.29$ &  - & - \\
J14444444-5953009 & 221.185170 & -59.883598 & $16.17\pm0.10$ & $14.74\pm0.08$ &  $13.84\pm0.08$ &   G316.7049-00.0767 & $12.74\pm0.18$ & $12.75\pm0.17$ & - & - \\
J14444590-5948273 & 221.191264 & -59.807587 & $14.96\pm0.06$ & $14.62\pm0.08$ &  $14.13\pm0.06$ &   G316.7398-00.0091 & $13.87\pm0.14$ & $13.53\pm0.33$ & - & - \\
J14444604-5951584 & 221.191872 & -59.866234 & $15.29\pm0.08$ & $14.62\pm0.10$ &  $14.10\pm0.12$ &   G316.7153-00.0624 & $14.01\pm0.13$ & $14.19\pm0.33$ & - & - \\
J14444694-5946520 & 221.195624 & -59.781132 & $15.69\pm0.07$ & $15.42\pm0.12$ &  $15.00\pm0.14$ &   G316.7530+00.0140 & $14.58\pm0.13$ & $14.63\pm0.22$ &  - & - \\
J14444710-5951048 & 221.196286 & -59.851341 & $13.25\pm0.03$ & $11.90\pm0.04$ &  $10.80\pm0.03$ &   G316.7236-00.0498 & $9.49\pm0.17$ & $9.28\pm0.07$ & $8.75\pm0.05$ & $8.36\pm0.09$ \\
J14444776-5953020 & 221.199010 & -59.883911 & $15.90\pm0.10$ & $15.28\pm0.12$ &
 $14.54\pm0.14$ &   G316.7111-00.0799 & $14.59\pm0.15$ & $14.27\pm0.29$ &  - & - \\
J14444821-5950155 & 221.200897 & -59.837666 & $13.56\pm0.03$ & $13.10\pm0.03$ &  $12.46\pm0.04$ &   G316.7314-00.0384 & $11.43\pm0.10$ & $11.42\pm0.10$ & - & - \\
J14444862-5946415 & 221.202618 & -59.778206 & $15.39\pm0.05$ & $14.76\pm0.08$ &  $14.29\pm0.08$ &   G316.7574+00.0150 & $13.29\pm0.08$ & $13.11\pm0.18$ &  - & - \\
J14444995-5945117 & 221.208153 & -59.753262 & $16.55\pm0.15$ & $15.27\pm0.12$ &  $14.29\pm0.09$ &   G316.7705+00.0366 & $13.27\pm0.08$ & $13.35\pm0.15$ & - & - \\
J14445285-5951361 & 221.220240 & -59.860054 & $14.34\pm0.02$ & $13.93\pm0.04$ &  $13.52\pm0.04$ &   G316.7308-00.0628 & $13.26\pm0.09$ & $13.32\pm0.11$ & - & - \\
J14445341-5946501 & 221.222550 & -59.780602 & $15.20\pm0.05$ & $13.31\pm0.04$ &  $11.94\pm0.04$ &   G316.7655+00.0087 & $10.27\pm0.08$ & $9.84\pm0.07$ & $9.63\pm0.05$ & $8.78\pm0.05$ \\
J14445354-5945305 & 221.223102 & -59.758499 & $15.94\pm0.10$ & $15.16\pm0.11$ &  $14.64\pm0.12$ &   G316.7751+00.0286 & $14.21\pm0.12$ & - & - & - \\
J14445482-5948430 & 221.228429 & -59.811954 & $15.88\pm0.08$ & $14.81\pm0.08$ &  $13.73\pm0.06$ &   G316.7549-00.0209 & $12.10\pm0.09$ & $11.90\pm0.21$ & - & - \\
J14445482-5952244 & 221.228439 & -59.873451 & $14.15\pm0.04$ & $13.54\pm0.13$ &  $13.07\pm0.30$ &   G316.7289-00.0767 & $13.66\pm0.16$ & $13.45\pm0.26$ & - & - \\
J14445891-5949593 & 221.245472 & -59.833141 & $15.56\pm0.07$ & $14.87\pm0.10$ &  $14.21\pm0.10$ &   G316.7538-00.0437 & - & - & $9.43\pm0.22$ & - \\
J14445959-5949534 & 221.248297 & -59.831509 & $16.33\pm0.13$ & $15.22\pm0.12$ &  $14.10\pm0.09$ &   G316.7557-00.0429 & $13.49\pm0.19$ & - & - & - \\
J14450134-5952384 & 221.255589 & -59.877350 & $15.25\pm0.07$ & $14.75\pm0.06$ &  $14.32\pm0.11$ &   G316.7396-00.0861 & $13.84\pm0.15$ & - & - & - \\
J14450302-5949357 & 221.262599 & -59.826584 & $15.69\pm0.15$ & $14.48\pm0.17$ &  $13.50\pm0.17$ &   G316.7642-00.0414 & $12.03\pm0.13$ & $12.00\pm0.19$ & - & - \\
J14450389-5951202 & 221.266232 & -59.855637 & $15.46\pm0.06$ & $14.93\pm0.10$ &  $14.42\pm0.10$ &   G316.7537-00.0686 & $13.72\pm0.17$ & - & - & - \\
J14450513-5945475 & 221.271415 & -59.763199 & $13.51\pm0.04$ & $13.14\pm0.05$ &
 $12.79\pm0.06$ &   G316.7951+00.0140 & $12.97\pm0.12$ & $13.13\pm0.22$ & - & - \\
J14450519-5949046 & 221.271647 & -59.817947 & $14.85\pm0.07$ & $13.12\pm0.05$ &  $11.94\pm0.05$ &   G316.7720-00.0356 & $10.26\pm0.05$ & $9.71\pm0.10$ & $9.25\pm0.09$ & - \\ 
J14450543-5952088 & 221.272625 & -59.869129 & $14.40\pm0.03$ & $14.11\pm0.04$ &  $13.89\pm0.05$ &   G316.7508-00.0821 & $13.54\pm0.15$ & $13.12\pm0.19$ & - & - \\
J14451464-5949435 & 221.311020 & -59.828758 & $16.04\pm0.14$ & $14.64\pm0.11$ &  $13.56\pm0.09$ &   G316.7854-00.0538 & $12.76\pm0.16$ & $12.46\pm0.16$ & - & -\\
J14451491-5950417 & 221.312136 & -59.844917 & $16.50\pm0.15$ & $15.20\pm0.13$ &  $14.05\pm0.07$ &   G316.7791-00.0687 & $12.68\pm0.07$ & $12.53\pm0.12$ & - & - \\
J14452048-5951525 & 221.335341 & -59.864597 & $16.22\pm0.13$ & $13.85\pm0.06$ &  $12.11\pm0.03$ &   G316.7813-00.0914 & $10.18\pm0.34$ & $9.77\pm0.16$ & $8.92\pm0.35$ & - \\
J14452136-5946106 & 221.339000 & -59.769630 & $16.36\pm0.15$ & $15.10\pm0.10$ &  $14.12\pm0.07$ &   G316.8232-00.0062 & $13.38\pm0.12$ & $13.36\pm0.35$ & - & - \\
J14452297-5951192 & 221.345719 & -59.855347 & $14.82\pm0.04$ & $14.22\pm0.05$ &  $13.70\pm0.04$ &   G316.7899-00.0852 & $12.04\pm0.09$ & $11.92\pm0.11$ & - & - \\
J14453074-5946116 & 221.378095 & -59.769901 & $15.28\pm0.05$ & $14.68\pm0.08$ &  $14.14\pm0.08$ &   G316.8410-00.0149 & $13.07\pm0.08$ & $13.17\pm0.21$ & - & - \\
J14453230-5945395 & 221.384600 & -59.760975 & $5.47\pm0.02$ & $4.50\pm0.08$ & $3.88\pm0.25$ &   G316.8477-00.0081 & - & $4.22\pm0.08$ & $3.75\pm0.02$ & $4.07\pm0.06$ \\
J14454091-5948089 & 221.420467 & -59.802486 & $16.21\pm0.15$ & $14.49\pm0.09$ &  $13.41\pm0.07$ &   G316.8464-00.0534 & $12.75\pm0.08$ & $12.67\pm0.13$ &  $11.72\pm0.24$ & - \\
J14454329-5952322 & 221.430399 & -59.875637 & $14.85\pm0.04$ & $14.30\pm0.06$ &  $13.92\pm0.06$ &   G316.8198-00.1217 & $13.96\pm0.19$ & - & - & - \\
J14454395-5950207 & 221.433157 & -59.839100 & $15.25\pm0.05$ & $13.97\pm0.06$ &  $12.58\pm0.04$ &   G316.8365-00.0892 & $11.15\pm0.08$ & $11.05\pm0.11$ & - & - \\
J14454572-5945582 & 221.440525 & -59.766190 & $14.01\pm0.07$ & $13.57\pm0.06$ &  $13.22\pm0.08$ &   G316.8709-00.0248 & $13.33\pm0.08$ & $12.84\pm0.18$ & - & - \\
\hline
\end{supertabular}
}
\end{center}
%\end{table}
\end{landscape}                                                                                 

\clearpage

\clearpage
\appendix

\section{Electron density and effective temperature of ionising radiation from ISO LWS spectrum}
                                                                                
 The ratio of flux in [O III] 52 $\mu$m to that in
[O III] 88 $\mu$m (hereafter denoted as [O III] 52/88) can be used to estimate
the electron density, $n_e$, in this region (Rubin et al., \cite{Ru94}). This is
because these are atomic fine structure lines of the same ionic species and
are emitted from levels with nearly the same excitation temperature. The ratio
[O III] 52/88 $\sim 1.4$ (see Table \ref{iso_tab1}) implies $n_e\sim300$ 
cm$^{-3}$ from the semi-empirical treatment of Rubin et al., (\cite{Ru94}).
                                                                                
It is also possible to estimate the effective
temperature of the ionising radiation ($T_{eff}$) using the fluxes in
the [N III] 57 $\mu$m and [N II] 122 $\mu$m lines under the assumption that
the nebula is ionisation bounded (Rubin et al., \cite{Ru94}). The volume
emissivities of [N III] 57 $\mu$m and [N II] 122 $\mu$m were obtained for
$n_e$ determined above. It is possible to derive the ion
abundance ratio N$^{++}$/N$^{+}$ using these volume emissivities and fluxes
 in the lines (Figure 4 of Rubin et al., \cite{Ru94}). This abundance
ratio is found to be N$^{++}$/N$^{+}$ = 0.91 which corresponds to an effective
temperature of $\sim37,500$ K.


\begin{thebibliography}{}
\bibitem[2004]{Al04} Allen, L. E., Calvet, N., D'Alessio, P. et al. 2004, ApJS, 154, 363
\bibitem[1990]{Au90} Aumann, H. H., Fowler, J. W. \& Melnyk, M., 1990, AJ, 99,
1674
\bibitem[2003]{Be03} Benjamin, R. A., Churchwell, E., Babler, B. L, et al. 
2003, PASP, 115, 953
\bibitem[1988]{Be88} Bessel, M. S. \& Brett, J. M., 1988, PASP, 100, 1134
\bibitem[1999]{Bo99} Bock, D. C., Large, M. I. \& Sadler, E. M. 1999, AJ, 117, 1578
\bibitem[1996]{Br96} Bronfman, L., Nyman, L.-A. \& May, J., 1996, A\&AS, 115, 81
\bibitem[2006]{Bu06} Busfield, A. L., Purcell, C. R., Hoare, M. G., Lumsden, S.
 L., Moore, T. J. T. \& Oudmaijer, R. D., 2006, MNRAS, 366, 1096
\bibitem[1987]{Ca87} Caswell, J. L. \& Haynes, R. F., 1987, AuJPh, 40, 215 
\bibitem[1989]{Ca89} Caswell, J. L., Batchelor, R. A., Forster, J. R. \& 
Wellington, K. J., 1989, AuJPh, 42, 331
\bibitem[1995]{Ca95} Caswell, J. L., Vaile, R. A., Ellingsen, S. P., Whiteoak, 
J. B. \&  Norris, R. P., 1995, MNRAS, 272, 96
\bibitem[1996]{Cl96} Clegg, P. E., Ade, P. A. R., Armund, C. et al. 1996, A\&A,
 315, L38
\bibitem[1984]{Dr84} Draine, B. T. \& Lee, H. M., 1984, ApJ, 285, 89
\bibitem[2004]{Fa04} Fazio, G. G., Hora, J. L., Allen, L. E. et al. 2004, ApJS, 154, 10
\bibitem[1996]{Fe96} Ferland, G. J., 1996, \textit{Hazy}, a brief introduction to CLOUDY, Univ. of Kentucky, Dept. of Phys. and Astron. Internal Reports.
\bibitem[1984]{Ga84} Gardner, F. F. \&  Whiteoak, J. B., 1984, MNRAS, 210, 23
\bibitem[1988]{Gh88} Ghosh, S. K., Iyengar, K. V. K., Rengarajan, T. N., Tandon, S. N., Verma, R. P. \& Daniel, R. R., 1988, ApJ, 330, 928
\bibitem[2002]{Gh02} Ghosh, S. K. \& Ojha, D. K., 2002, A\&A, 288, 326
Cernicharo, J., 2004, ApJ, 600, 214
\bibitem[1978]{Gu78} Gull, S. F. \& Daniell, G. J., 1978, Nature, 272, 686
\bibitem[1983]{Hi83} Hildebrand, R. H., 1983, QJRAS, 24, 267
\bibitem[1999]{Hu99} Huang, M, Bania, T. M., Bolatto, A. et al. 1999, ApJ, 517, 282
\bibitem[1996]{Ju96} Juvela, M., 1996, A\&AS, 118, 191
\bibitem[1992]{La92} Lada, C. J. \& Adams, F. C., 1992, ApJ, 393, 278
\bibitem[1993]{La93} Laor, A. \& Draine, B. T., 1993, ApJ, 402, 441
\bibitem[2003]{Ll03} Lloyd, C., Lerate, M. R. \& Grundy, T. W., 2003, `Uniformly
 Processed LWS L01 Spectra', ISO Technical Note 17 (Madrid: ISO Data Centre)
\bibitem[1997]{Me97} Meyer, M. R., Calvet, N. \& Hillenbrand, L. A., 1997, AJ, 
114, 288
\bibitem[1982]{Me82} Mezger, P. G., Mathis, J. S. \& Panagia, N., 1982, A\&A,
105, 372
\bibitem[2002]{Mi02} Mizutani, M., Onaka, T. \& Shibai, H., 2002, A\&A, 382, 610
\bibitem[1999]{Mo99} Mookerjea, B. \& Ghosh, S. K., 1999, J. Astrophys. Astr.
20, 1
\bibitem[2000]{Mo00} Mookerjea, B., Ghosh, S. K., Rengarajan, T. N., Tandon, S. N. \& Verma, R. P., 2000, AJ, 120, 1954
\bibitem[2002]{Mo02} Morisset, C., Schaerer, D., Mart\'in-Hern\'andez, N. L., et al. 2002, A\&A, 386, 558 
\bibitem[2005]{Pe05} Peeters, E., Mart\'in-Hern\'andez, N. L., Rodr\'iguez-Fern\'andez, N. J. \& Tielens, X., 2005, \textit{Space Science Review ISO Special Issue}, Springer
\bibitem[2001]{Pr01} Price, S. D., Egan, M. P., Carey, S. J.,
Mizuno, D. R., \& Kuchar, T. A., 2001, AJ, 121, 2819
\bibitem[1985]{Ri85} Rieke, G. H. \& Lebofsky, M. J., 1985, ApJ, 288, 618
\bibitem[1994]{Ru94} Rubin, R. H., Simpson, J. P., Lord, S. D., Colgan, S. W. J., Erickson, E. F. \& Haas, M. R., 1994, ApJ, 420, 772
\bibitem[2000]{Sa00} Sandell, G., 2000, A\&A, 358, 242
\bibitem[1976]{Sc76} Scoville, N. Z. \& Kwan, J. 1976, ApJ, 206, 718
\bibitem[2000]{Vi00} Vilas-Boas, J. W. S. \& Abraham, Z., 2000, A\&A, 355, 1115
\bibitem[1998]{Wa98} Walsh, A. J., Burton, M. G., Hyland, A. R. \& Robinson, G., 1998, MNRAS, 301, 640
\bibitem[2004]{We04} Werner, M. W., Roellig, T. L., Low, F. J., et al. 2004, 
ApJS, 154, 1
\bibitem[1983]{Wh83} White, G. J. \& Phillips, J. P., 1983, MNRAS, 202, 255
\bibitem[1982]{Wh82} Whiteoak, J. B., Otrupcek, R. E. \& Rennie, C. J., 1982, 
PASAu, 4, 434
\end{thebibliography}
\end{document}